\documentclass[11pt]{elsarticle}
\usepackage{amsmath}
\usepackage{amssymb}
\usepackage{verbatim}
\usepackage{caption}
\usepackage{algorithm2e}
\usepackage{listings}
\usepackage{textcomp}
\usepackage[usenames,dvipsnames]{xcolor}
\usepackage{graphicx}
\usepackage[textfont=up]{caption}
\usepackage{float}
\usepackage{color}
\usepackage{tikz}
\usetikzlibrary{arrows,shapes,scopes}
\usetikzlibrary{calc}
\usetikzlibrary{graphs}
\usepackage{caption}
\usepackage{subcaption}
\usepackage{hyperref}
\hypersetup{colorlinks=true,linkcolor=blue,citecolor=purple}
\graphicspath{{./pictures/}}
\newsavebox{\bigleftbox}

\usepackage{listings}
\definecolor{cmtgray}{gray}{0.45}
\definecolor{cmtmgray}{gray}{0.20}
\lstdefinestyle{customc}{
  belowcaptionskip=.3\baselineskip,
  escapeinside={/*@}{@*/},
  breaklines=true,
  frame=tB,
  xleftmargin=\parindent,
  language=C++,
  showstringspaces=false,
  basicstyle=\ttfamily,
  keywordstyle=\bfseries\color{green!40!black}\textbf,
  commentstyle=\color{cmtgray}\ttfamily,
  identifierstyle=\color{black!90!white},
  stringstyle=\color{black},
  captionpos=b,
  linewidth=\textwidth,
 morekeywords={}
}
\newcommand{\ductteipw}{DuctTeip}
\newcommand{\ductteip}{DuctTeip~}
\newcommand{\superglue}{SuperGlue~}
\newcommand{\supergluew}{SuperGlue}

\begin{document}
\begin{frontmatter}

\title{\ductteipw: An efficient programming model for distributed task based parallel computing\tnoteref{t1}}
\tnotetext[t1]{This work was supported in part by the Swedish Research Council and carried out within the Linnaeus centre of excellence UPMARC, Uppsala Programming for Multicore Architectures Research Center.}

\author[1]{Afshin Zafari}
\ead{afshinzafari@gmail.com}

\author[1]{Elisabeth Larsson\corref{c1}}
\ead{elisabeth.larsson@it.uu.se}
\cortext[c1]{Corresponding author}

\author[1]{Martin Tillenius}
\ead{martin.tillenius@gmail.com}

\address[1]{Uppsala University, Department of Information Technology, Box~337, SE-751~05 Uppsala, Sweden}

\begin{abstract}
Current high-performance computer systems used for scientific computing typically combine shared memory computational nodes in a distributed memory environment. Extracting high performance from these complex systems requires tailored approaches. Task based parallel programming has been successful both in simplifying the programming and in exploiting the available hardware parallelism for shared memory systems. In this paper we focus on how to extend task parallel programming to distributed memory systems. We use a hierarchical decomposition of tasks and data in order to accommodate the different levels of hardware. We test the proposed programming model on two different applications, a Cholesky factorization, and a solver for the Shallow Water Equations. We also compare the performance of our implementation with that of other frameworks for distributed task parallel programming, and show that it is competitive.
\end{abstract}

\end{frontmatter}


\section{Introduction}\label{sec:intro}
Task based parallel programming models~\cite{JSFI90} have, over the last decade, been developed to meet the contemporary challenge of programming for rapidly evolving multicore hardware. 
In this programming paradigm, the work for the programmer is reduced to expressing an algorithm in terms of tasks and their accesses to shared data, while scheduling and execution of tasks in parallel are managed by an application independent run-time system. The run-time system is also responsible for ensuring that data dependencies are respected. The dependencies can, e.g., be deduced by the run-time system from user annotations of data accesses. 

While moving away from optimized software for a specific architecture, the role of the run-time system is to extract as high performance as possible from a wide range of multicore CPUs. Task based programming models provide some of the key components of multicore performance. Task sizes can be selected (when the application allows it) such that their data fits into the lowest level cache for efficient use of the memory hierarchy. Fine grained tasks also provide a high level of parallelism that can supply high numbers of threads with work. Local synchronization points between tasks (as opposed to global barriers), allow for asynchronous task execution, which can compensate for performance variations between threads due to, e.g., variable task sizes, resource contention, non-uniform memory access (NUMA) effects, or interference from simultaneously running processes in the kernel, run-time or other applications. 

A large number of frameworks for shared memory task parallel programming that address the common issues of portability of software, ease of programming, and performance on multicore and possibly heterogeneous computer systems have been developed. The type of parallelism that is provided varies from purely fork-join as in Cilk~\cite{BJKLRZ95} and Wool~\cite{Faxen09}, fork-join that allows data dependencies between sibling tasks, as implemented in Swan~\cite{VaTzeNi11},  OpenMP~\cite{OpenMP} version 4.0 or higher, and XKaapi~\cite{Virouleau2016}, to directed acyclic graph (DAG) based, where tasks are organized as nodes in a graph, where edges represent must-execute-before dependencies, as in Intel Threading Building Blocks~\cite{tbb}, OmpSs~\cite{DuranABLMMP11}, and StarPU~\cite{ATNW11}.

In this paper, we develop a distributed task programming model, and implement it in the distributed task parallel framework \ductteip. We use a hybrid parallel implementation with MPI for the distributed layer and Pthreads for the shared memory layer. We build on the shared memory library SuperGlue\footnote{\url{https://github.com/tillenius/superglue}}~\cite{Tillenius15}, developed by one of the authors. The reason for this choice is based on both the performance of SuperGlue and the suitability of the programming model for an extension to distributed memory.
 
In~\cite{Tillenius15}, SuperGlue was shown to perform better than ten of the other shared memory frameworks available at that time. 
A key component for the performance is the dependency management through data versioning. The resulting programming model is very similar to the DAG based models, but has some additional flexibility regarding dependencies.  

\ductteip has been developed, not primarily to compete with other frameworks, but to perform research on different ways to promote both flexibility and performance in the distributed setting. Understanding which trade-offs can be made and what their consequences are is relevant for the community that develops task parallel run-times, and accordingly, in the long term perspective, also for the users of the frameworks.

When we evaluate the performance of \ductteipw, we compare with the three major task frameworks/run-times with support for distributed memory architectures that implement a similar programming model: StarPU~\cite{AAFNT12}, Cluster OmpSS~\cite{TFGBAL11}, and PaRSEC~\cite{BBDFHD13}. We introduce each in some detail below.

The MPI version of StarPU~\cite{AAFNT12} maps the task DAG to the computational nodes. The edges in the DAG (dependencies) that cross the boundary between two computational nodes are mapped to MPI-based data transfer routines. Any MPI process can create and submit tasks, and the global task graph is known to all. The communication calls are initiated at task submission time, which may lead to high memory consumption if too many receive buffers are allocated in advance. This issue is investigated and handled through pausing task submission in~\cite{sergent16}. 

Cluster OmpSs~\cite{TFGBAL11} employs centralized task creation and submission with one master process that submits tasks to all other processes. A task can generate child tasks at execution time. These child tasks can have dependencies to  their sibling tasks, but not to other tasks.

PaRSEC~\cite{BBDFHD13,BBDHLD12} provides a highly efficient platform for distributed task based parallel execution. An annotated user code is here first (automatically) translated into a parametrized DAG representation. In particular, this representation can generate information about the predecessors and successors of each task. There is no need to store or expand the full graph, instead task specific information can be examined locally through queries on the DAG representation. Scheduling and task queues are fully NUMA aware and exploit the memory hierarchy to promote data locality. Communication is asynchronous and managed by a dedicated thread on each node.

In the \ductteip programming model, we do not assume any pre-knowledge regarding the task graph. Tasks can be generated dynamically, and tasks can be submitted by other tasks. Dependencies are of the DAG type, but with an extension to commutative dependencies through the data versioning model. Dependencies between tasks that are children of other tasks are not limited to siblings, but can be to other tasks on that level, which will be described in more detail later. Furthermore, the data that tasks work on can in principle have any format as protection of shared data is managed indirectly through data handles. An early description of the \ductteip programming model is available in~\cite{ZaTiLa12}. 
In this paper, we develop and evaluate two main ideas within \ductteip for distributed task parallel programming
\begin{itemize}
\item a hierarchical task and data model based on data versioning, which has several benefits regarding flexibility, utilization of the memory hierarchy, and network loading,
\item an efficient communication model, which avoids redundant messaging, and enables task and data migration (for future work on dynamic load balancing).  
\end{itemize}

Distributed task frameworks that are based on, to some extent, other programming models are, e.g., Charm++~\cite{Kale11,KaKri93}, which implements message-driven execution, where the arrival of a data object prompts the execution of a task, and Chunks and Tasks~\cite{RubRud14}, where chunks are immutable data objects that can be split hierarchically into smaller chunks, and tasks are objects that work on chunks, and can generate new tasks.

The paper is organized in the following way: In Section~\ref{sec:versions} we briefly introduce dependency management through data versioning. Sections~\ref{sec:hier} and~\ref{sec:comm} describe the task and data model, and the communication model. In Section~\ref{sec:user} we discuss how to support programmers in using the \ductteip model efficiently. In Section~\ref{sec:experiments} the performance and scalability of \ductteip are investigated and compared to that of StarPU, Cluster OmpSs, PaRSEC, and also ScaLAPACK.
Finally, conclusions based on the results are made in Section~\ref{sec:conc}.

\section{Tracking dependencies through data versioning}\label{sec:versions}
A detailed description of the \superglue programming model including data versioning can be found in~\cite{Tillenius15}. Here we give brief account to serve as a basis for the discussion in the following sections of the distributed case.

In this model, shared data is protected by a construct that we call a \textit{handle}. Since the handle is not physically connected to a memory location, it provides flexibility to the programmer in the sense that we can protect any type of shared data or shared resource~\cite{Tillenius15,TiLaBaMa15}.  

Each data handle is equipped with a version counter, which allows the run-time system to order data accesses (at execution time) such that data dependencies are respected. The versioning system is initialized at task submission time based on the assumption that tasks are submitted in a consistent sequential order.

At task submission time, the run-time system counts the (future) accesses to each data handle, and for each task, records the \textit{required version number} of each of the accessed data handles. How the required version is incremented depends on the type of access. Tasks that require the same version number can be reordered.

During execution, the \textit{run-time version number} of each handle is incremented after each access. Technically, this operation is performed at the completion of the execution of a task. In this way, the run-time system can compare the required version numbers for a certain task with the current run-time version numbers of the involved handles in order to determine if the task is ready to run.

\begin{figure}[!htb]
\centering
\hfill
\includegraphics[width=0.2\textwidth]{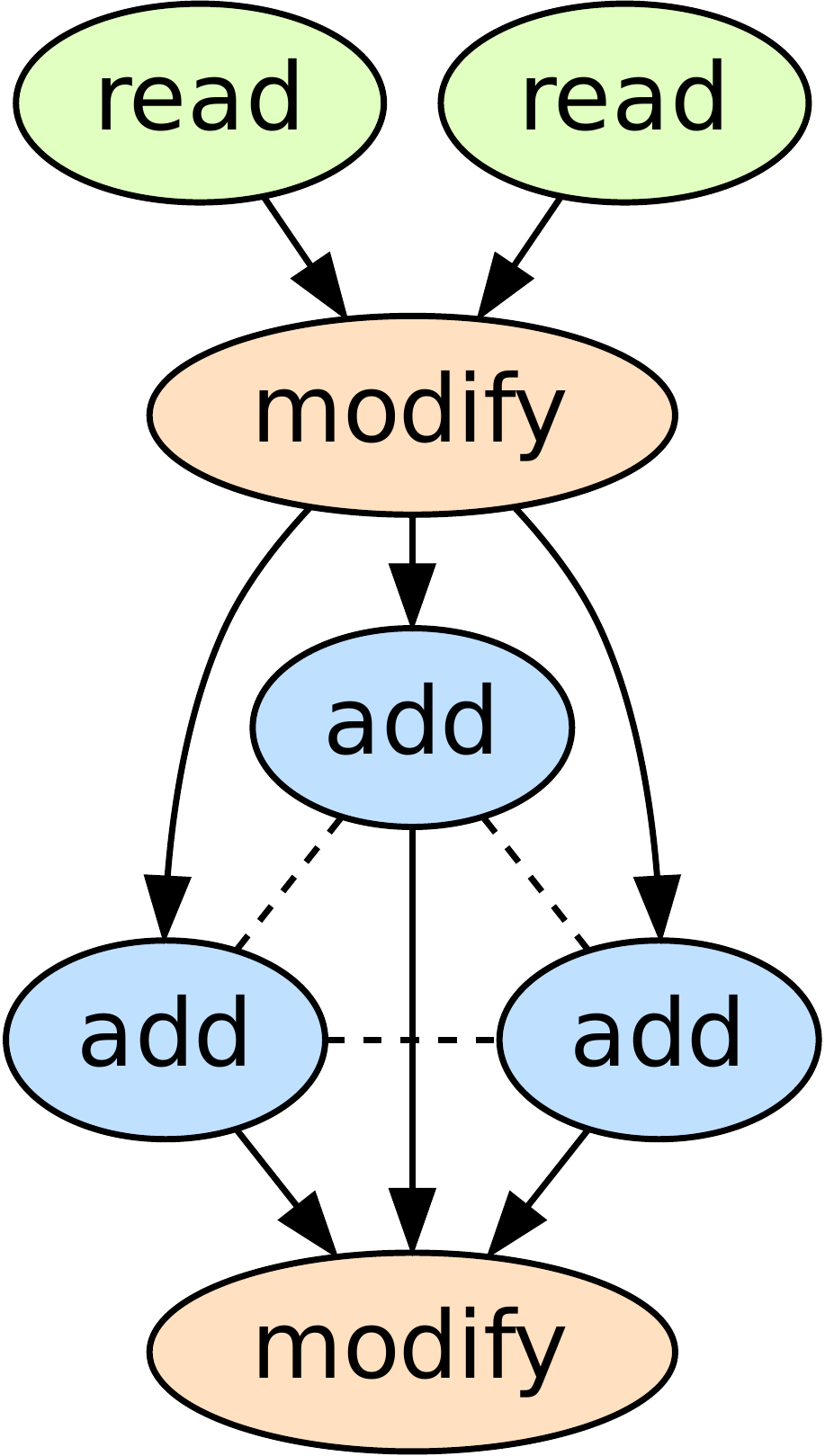}\hfill
\includegraphics[width=0.2\textwidth]{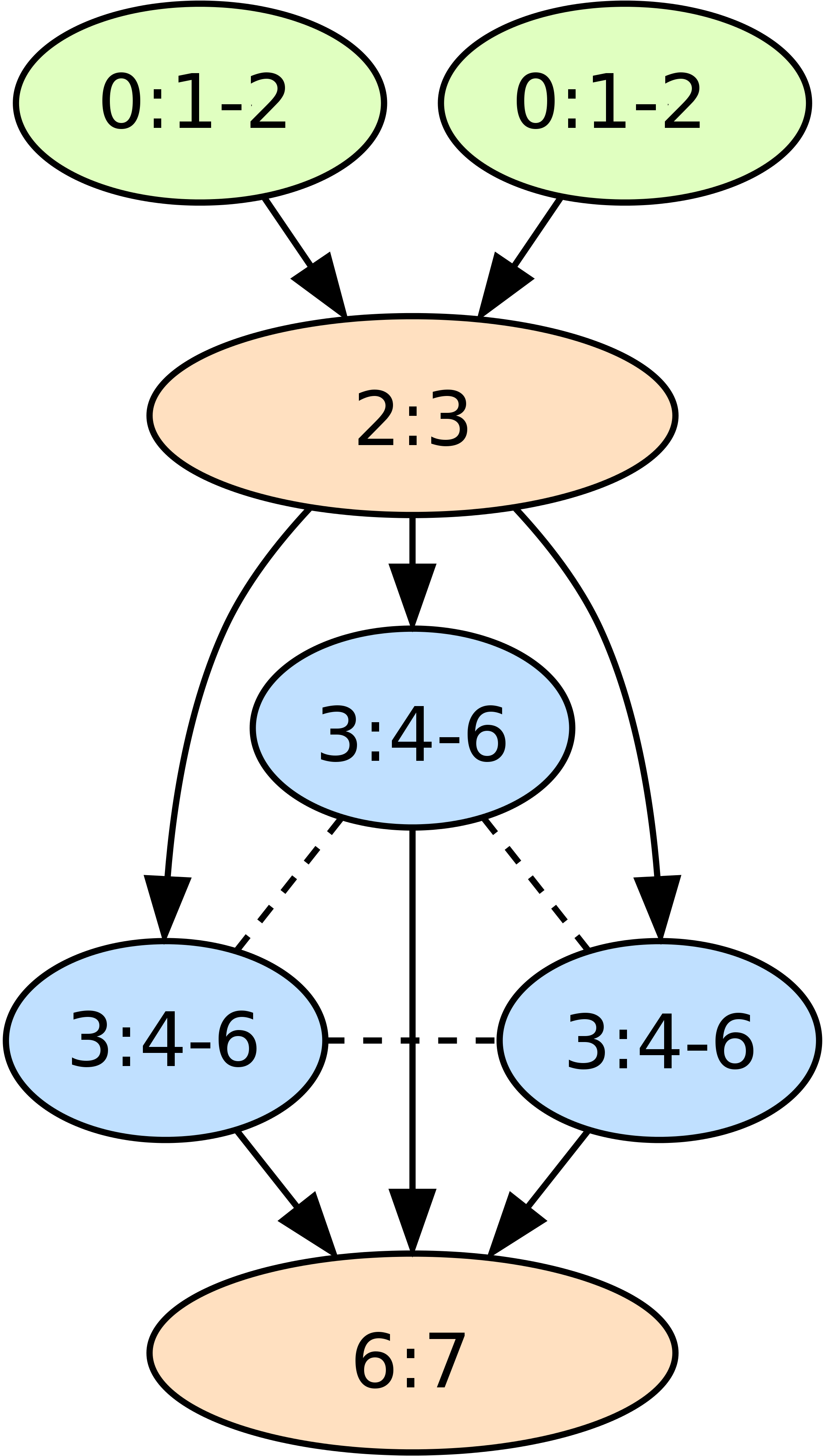}
\hfill\hspace{0mm}
\caption{Left: A small task graph where all accesses (the type is indicated for each task) are assumed to be to the same shared data. Dashed lines indicate a commutative dependency (any order, but one at a time). Right: The version $v$ required by the task to run, and the run-time version $r$ produced by the task given in the format $v:r$. A range of numbers indicates that any number in the range can be realized in an execution.}
\label{fig:dep}
\end{figure}

Figure~\ref{fig:dep} provides a small example of a task graph and the corresponding versions. The version of each handle starts at 0. The two read tasks can run immediately and in any order, but the first task that modifies the data must wait until both reads have finished. Hence, the modify task requires version 2. The three adds can run in any order, and when all of them are finished, the final modify task requiring version 6 is allowed run.


When the arguments of a task are examined to determine if the task is ready to run, and an argument that is not ready is encountered, the task is placed in a queue for that particular data version. This means that as soon as the data is ready, the run-time system knows which tasks to wake up either for execution or for examining of the next data in the argument list. In this way, the need for traversing long lists of waiting tasks is reduced, and the run-time can deploy the ready tasks at the thread where the data is locally cached. This improves the data locality of the execution, and also reduces the need for a particular scheduling strategy. In \superglue load balancing is achieved dynamically through task stealing. 

Scheduling to improve data locality and minimizing NUMA effects in shared memory task parallel execution is an active research area~\cite{Virouleau2016,Ceballos17,Ceballos15,DHDPC14,MJB15,PanPai15,PATM14,YLTM13} and can also be coupled to energy considerations~\cite{jimborean14,KBSK13}.

\section{The hierarchical \ductteip task and data model}\label{sec:hier}
We exploit a hierarchical task and data model to extract performance from the hardware in the distributed task parallel case.
%
%
The corresponding hierarchical target architecture is a distributed memory system (cluster) of shared memory computational nodes. A key idea in this approach is to use large data partitions for communication and small data partitions for computations. Communicating a single large data block allows for reducing the overhead of the communication compared with multiple smaller messages. Then locally,  further decomposing the large data into smaller parts increases the cache locality of the data in the local computations. 

To make the description of the hierarchical model general, we introduce the concept of levels, where level 0 is the top level (root), and level L is the finest level (leaves). The typical data at level 0 consists (symbolically) of the unpartitioned data structure and tasks submitted at level $\ell$ works on level $\ell+1$ data. We also allow for data to be allocated and created at levels $\ell>0$.

\begin{figure}[!htb]
\centering
\begin{tikzpicture}
        \begin{scope}
          \node at (.5,1.5) [anchor =north] {\small{Data}};
          \draw [thick,fill=blue] (0,0) rectangle (1.1,1) ;
          \node at ( 2.5,.5) [ anchor = east] {level 0};
          \node at ( 5,.5) [ anchor = west] {Task submission};
          \draw[-latex,thick] (6.4,.3)--(6.4,-.5);
          \draw [thin,fill=purple] (3.5,.5) circle (.5cm and 0.25cm);
          \node at ( 3.5,1.2) [ anchor = north] {\small{Algorithm}};
          \draw [thick,fill=blue] (0,-.7) rectangle ( .5,-.2);
          \draw [thick,fill=blue] (0.6,-.7) rectangle ( 1.1,-.2);
          \draw [thick,fill=blue] (0,-1.3) rectangle ( .5,-.8);
          \draw [thick,fill=blue] (0.6,-1.3) rectangle ( 1.1,-.8);
          \node at ( 2.5,-.7) [ anchor = east] {level 1};
          \node at ( 5,-0.7) [ anchor = west] {DuctTeip tasks};
          \draw[-latex,thick] (6.4,-.9)--(6.4,-1.6);
          \node at (4.0, -.15) [anchor=east] {\small{Tasks}};
          \draw [thin,fill=purple] (3.0,-.5) circle (0.25cm and  0.125cm);
          \draw [thin,fill=purple] (3.6,-.5) circle (0.25cm and  0.125cm);
          \draw [thin,fill=purple] (4.2,-.5) circle (0.25cm and  0.125cm);
          
          \draw [thin,fill=purple] (3.3,-.9) circle (0.25cm and  0.125cm);
          \draw [thin,fill=purple] (3.9,-.9) circle (0.25cm and  0.125cm);
		
          \draw [thick,fill=blue] (0.25,-1.75) rectangle (.5,-1.5);
          \draw [thick,fill=blue] (0.55,-1.75) rectangle (.8,-1.5);
          \draw [thick,fill=blue] (0.25,-2.05) rectangle (.5,-1.8);
          \draw [thick,fill=blue] (0.55,-2.05) rectangle (.8,-1.8);
          \node at ( 2.5,-1.8) [ anchor = east] {level 2};
          \node at ( 5,-1.8) [ anchor = west] {SuperGlue tasks};

          \draw [thin,fill=purple] (3.0,-1.6) circle (0.125cm and  0.0625cm);
          \draw [thin,fill=purple] (3.3,-1.6) circle (0.125cm and  0.0625cm);
          \draw [thin,fill=purple] (3.6,-1.6) circle (0.125cm and  0.0625cm);
          \draw [thin,fill=purple] (3.9,-1.6) circle (0.125cm and  0.0625cm);		
          \draw [thin,fill=purple] (4.2,-1.6) circle (0.125cm and  0.0625cm);
    
          \draw [thin,fill=purple] (3.15,-1.9) circle (0.125cm and  0.0625cm);
          \draw [thin,fill=purple] (3.45,-1.9) circle (0.125cm and  0.0625cm);
          \draw [thin,fill=purple] (3.75,-1.9) circle (0.125cm and  0.0625cm);
          \draw [thin,fill=purple] (4.05,-1.9) circle (0.125cm and  0.0625cm);
        \end{scope}
      \end{tikzpicture}
\caption{Schematic illustration of hierarchical data and tasks in the currently implemented three level model.}\label{fig:hier}
\end{figure}
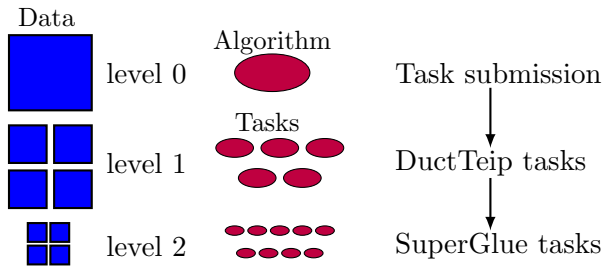

We exemplify with a basic three level configuration, where the distributed memory framework \ductteip is used at level 0 (working on level 1 data), and the shared memory library \superglue is employed at level 1 (working on level 2 data). For an illustration, see Figure~\ref{fig:hier}.
In the discussion of data and tasks below, we describe the general case with L data levels.

In the following subsections, we explain the hierarchical approach for defining data and tasks and how it affects the dependency tracking and the task execution in the context of distributed parallel programming.

\subsection{The hierarchical data structures}\label{sec:hierdata}


Conceptually, we consider general data types such as dense matrices, sparse matrices, and tree structures. Examples are given in terms of dense matrices below to provide clear illustrations. For each type of data that can be represented, a parametrized splitting scheme needs to be provided. The parametrization allows for tuning partition sizes with respect to the numbers of nodes and cores as well as the local cache sizes. For a square matrix, a standard splitting is to divide a matrix block $A_{\ell-1}$ at level $\ell-1$ into $n_\ell\times n_\ell$ blocks $A_{\ell-1}(i,j)$, $i,j=1,\ldots,n_\ell$ at level $\ell$.

As indicated, we also need an indexing method associated with the splitting. The algorithm at level $\ell-1$ submits tasks formulated in terms of the data partitions $A_{\ell-1}(i_1,\ldots,i_d)$, where $d$ is the number of dimensions of the splitting and the indexing.

\begin{figure}[!htb]
\begin{subfigure}[t]{0.45\textwidth}
\centering

\begin{tikzpicture}

\draw [step=.7,very thick,color=black!20!white] (0,4.89) grid (2.1,6.3);
\node at (.35,5.9) {$P_{0}$};
\node at (1.05,5.9) {$P_{1}$};
\node at (1.75,5.9) {$P_{2}$};

\node at (.35,5.25) {$P_{3}$};
\node at (1.05,5.25) {$P_{4}$};
\node at (1.75,5.25) {$P_{5}$};

\begin{scope}[xshift=0cm]
\draw [step=.7,very thick] ($(0,3.499) +(2.799,0)$) grid ($(2.8,6.3)+(02.8,0)$);

\node at ($(-1.4,5.9)+(4.5,0)$) {$\color{black!50!white}P_{0}$};
\node at ($(-0.7,5.9)+(4.5,0)$) {$\color{black!50!white}P_{1}$};
\node at ($(0,5.9)+(4.5,0)$) {$\color{black!50!white}P_{2}$};
\node at ($(0.7,5.9)+(4.5,0)$) {$\color{black!50!white}P_{0}$};
\node at ($(-1.4,5.2)+(4.5,0)$) {$\color{black!50!white}P_{3}$};
\node at ($(-0.7,5.2)+(4.5,0)$) {$\color{black!50!white}P_{4}$};
\node at ($(0,5.2)+(4.5,0)$) {$\color{black!50!white}P_{5}$};
\node at ($(0.7,5.2)+(4.5,0)$) {$\color{black!50!white}P_{3}$};

\node at ($(-1.4,5.2)+(4.5,-0.7)$) {$\color{black!50!white}P_{0}$};
\node at ($(-0.7,5.2)+(4.5,-0.7)$) {$\color{black!50!white}P_{1}$};
\node at ($(0,5.2)+(4.5,-0.7)$) {$\color{black!50!white}P_{2}$};
\node at ($(0.7,5.2)+(4.5,-0.7)$) {$\color{black!50!white}P_{0}$};
\node at ($(-1.4,5.2)+(4.5,-1.4)$) {$\color{black!50!white}P_{3}$};
\node at ($(-0.7,5.2)+(4.5,-1.4)$) {$\color{black!50!white}P_{4}$};
\node at ($(0,5.2)+(4.5,-1.4)$) {$\color{black!50!white}P_{5}$};
\node at ($(0.7,5.2)+(4.5,-1.4)$) {$\color{black!50!white}P_{3}$};
\end{scope}
\end{tikzpicture}

\caption{Block-cyclic ownership.}
\label{fig:pqgridpxq}
\end{subfigure}
%
\hspace{5mm}
\begin{subfigure}[t]{0.45\textwidth}
\centering
\begin{tikzpicture}
\begin{scope}[xshift=0cm]
\draw [step=.7,very thick,color=black!20!white] (0,5.59) grid (2.1,6.3);
\node at (.35,5.9) {$P_{0}$};
\node at (1.05,5.9) {$P_{1}$};
\node at (1.75,5.9) {$P_{2}$};
\end{scope}

\draw [step=.7,very thick] ($(0,3.499) +(2.799,0)$) grid ($(2.8,6.3)+(2.8,0)$);

\node at ($(-1.4,5.9)+(4.5,0)$) {$\color{black!50!white}P_{0}$};
\node at ($(-0.7,5.9)+(4.5,0)$) {$\color{black!50!white}P_{1}$};
\node at ($(0,5.9)+(4.5,0)$) {$\color{black!50!white}P_{2}$};
\node at ($(0.7,5.9)+(4.5,0)$) {$\color{black!50!white}P_{0}$};
\node at ($(-1.4,5.2)+(4.5,0)$) {$\color{black!50!white}P_{0}$};
\node at ($(-0.7,5.2)+(4.5,0)$) {$\color{black!50!white}P_{1}$};
\node at ($(0,5.2)+(4.5,0)$) {$\color{black!50!white}P_{2}$};
\node at ($(0.7,5.2)+(4.5,0)$) {$\color{black!50!white}P_{0}$};

\node at ($(-1.4,5.2)+(4.5,-0.7)$) {$\color{black!50!white}P_{0}$};
\node at ($(-0.7,5.2)+(4.5,-0.7)$) {$\color{black!50!white}P_{1}$};
\node at ($(0,5.2)+(4.5,-0.7)$) {$\color{black!50!white}P_{2}$};
\node at ($(0.7,5.2)+(4.5,-0.7)$) {$\color{black!50!white}P_{0}$};
\node at ($(-1.4,5.2)+(4.5,-1.4)$) {$\color{black!50!white}P_{0}$};
\node at ($(-0.7,5.2)+(4.5,-1.4)$) {$\color{black!50!white}P_{1}$};
\node at ($(0,5.2)+(4.5,-1.4)$) {$\color{black!50!white}P_{2}$};
\node at ($(0.7,5.2)+(4.5,-1.4)$) {$\color{black!50!white}P_{0}$};

\end{tikzpicture}

\caption{Column-cyclic ownership.}
\label{fig:pqgrid1xq}
\end{subfigure}




\caption{Two different process grids $(p,q)=(2,3)$ and $(p,q)=(1,3)$ for the processes $P_i$, and the resulting block cyclic distributions for a $4\times 4$ block matrix.}
\label{fig:pqgrid}
\end{figure}
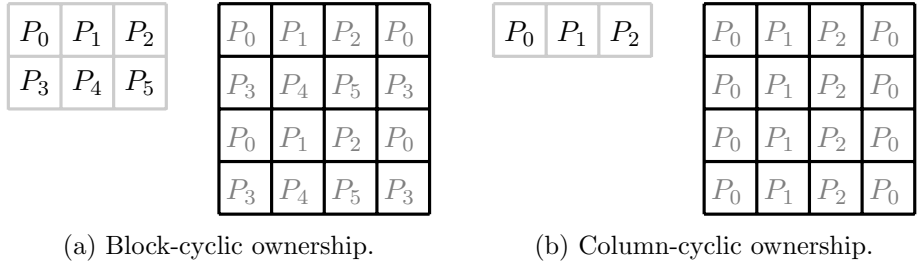

When a shared data structure $A_0$ within the parallel framework is initialized at level~0, we do not want to allocate the whole structure anywhere. The distributed level is in our case level $\ell=1$, and this is where the data is associated with a node and allocated locally. The data ownership is determined from a process grid topology and a distribution scheme. An example of how a block-cyclic distribution of a dense matrix interacts with the process grid is shown in Figure~\ref{fig:pqgrid}.



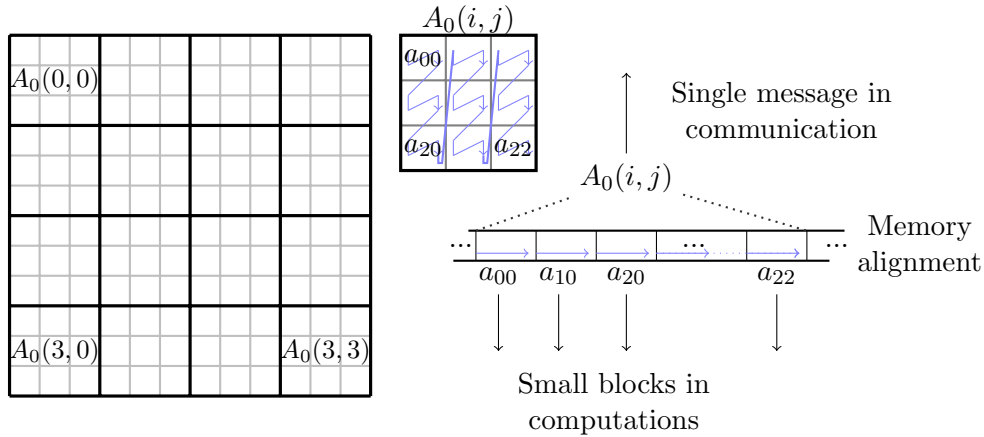
\begin{figure}
\centering
\begin{tikzpicture}

\draw [step=.4,color=gray!50!white,thick] (0,3.6) grid (4.8,8.4);
\draw [step=1.2,very thick] (0,3.599) grid (4.8,8.4);

\node at (.6,7.8) {\small $A_0(0,0)$};
\node at (.6,4.2) {\small $A_0(3,0)$};
\node at (4.2,4.2) {\small $A_0(3,3)$};

\begin{scope}[xshift=4cm,yshift=6.6cm]
\node at (2.1,2.) {$ A_0(i,j)$};
\draw [step=.6,thick,color=gray] (1.2,0) grid (3.0,1.8);
\draw [very thick] (1.2,0) rectangle ( 3.0,1.8);

\foreach \c in {0,1,2}{
	\foreach \r in {0,1,2}{
		\foreach \xf/\xt/\yf/\yt/\a in {1.3/1.7/1.4/1.6/0.6}{
			\draw[-,blue!50!white,thin] ($ (\xf,\yt) + \a*(\c,-\r) $) -- ($ (\xf,\yf) +\a*(\c,-\r) $) ;
			\draw[-,blue!50!white,thin] ($ (\xf,\yf) + \a*(\c,-\r) $) -- ($ (\xt,\yt) +\a*(\c,-\r) $) ;
			\draw[->,blue!50!white,thin] ($ (\xt,\yt) +\a*(\c,-\r) $)  -- ($ (\xt,\yf) +\a*(\c,-\r) $) ;
		}
	}
}
\foreach \c in {0,1,2}{
	\foreach \r in {0,1}{
		\foreach \xf/\xt/\yf/\yt/\a in {1.3/1.7/1.4/1.6/0.6}{
			\draw[-,blue!50!white,thin] ($ (\xt,\yf) +\a*(\c,-\r) $) -- ($ (\xf,\yt) + \a*(\c,-\r-1) $);
		}
	}
}
\foreach \c in {0,1}{
		\foreach \xf/\xt/\yf/\yt/\a in {1.3/1.7/1.4/1.6/0.6}{
			\draw[-,blue!50!white,thick] ($ (\xt,\yf) +\a*(\c,-2) $) -- ($ (\xt,\yf-.1) +\a*(\c,-2) $) -- ($ (\xt+0.05,\yf-0.1) +\a*(\c,-2) $) -- ($ (\xf,\yt) + \a*(\c+1,0) $);
		}
}

\node at (1.5,1.5) {${\small a_{00}}$};
\node at (1.5,.3) {${\small a_{20}}$};
\node at (2.7,.3)  {${\small a_{22}}$};
\end{scope}

\begin{scope}[xshift=1.5cm,yshift=1cm]
\draw[thick] (4.4,4.4) -- (9.4,4.4);
\draw[thick] (4.6,4.8) -- (9.6,4.8);
\node at (4.5,4.6) {...};
\node at (9.5,4.6) {...};

\node [text width =3cm,align=center] at (10.6,4.6) {Memory \\alignment};

\node (Aij) at (6.7,5.5) {$A_0(i,j)$};
\draw[->] (Aij) -- (6.7,6.9);
\node  at(10.4,5.9) [anchor=south east,text width=3cm,align=center] {Single message in \\
communication} ;

\node at (7.6,4.6) {...};

\foreach \c in {0,1,2,3,4.5,5.5}{
	\draw ($ (4.7,4.4) +\c*(0.8,0) $) -- ($ (4.7,4.8) +\c*(0.8,0) $);
}
\foreach \c in {0,1,2,3,4.5}{
	\draw[->,blue!50!white,thin] ($ (4.7,4.5) +\c*(0.8,0) $) -- ($ (5.4,4.5) +\c*(.8,0) $);
}
\draw[->,dotted,blue!50!white,thin] ($ (4.7,4.5) +3.*(0.8,0) $) -- ($ (5.4,4.5) +4.5*(.8,0) $);
\draw [gray!40!black,dotted,thick](4.7,4.8) -- (Aij)-- ($ (4.7,4.8) +5.5*(0.8,0) $);

\node (a00) at (5.0,4.2) {${\small a_{00}}$};
\node (a10) at (5.8,4.2) {${\small a_{10}}$};
\node (a20) at (6.7,4.2) {${\small a_{20}}$};
\node (a22) at (8.7,4.2) {${\small a_{22}}$};
\foreach \a in{ a00,a10,a20,a22}{
	\draw[->] (\a) -- ($ (\a) - (0,1.0)$);
}
\node (comp) at (4.9,2.5)  [right,text width =3.0cm,align=center] 
{Small blocks in \\
computations
};
\end{scope}
\end{tikzpicture}

\caption{An example of hierarchical data decomposition. A matrix is partitioned into 4$\times$4 large blocks at level 1, each of which contains 3$\times$3 small blocks at level 2. The elements of each block at each level are linearly indexed in contiguous memory, as indicated by the blue line traversing one sub block at a time. The notation $a_{mn}$ is used for partitions of the level 1 data.}
\label{fig:twoleveldata}
\end{figure}

For efficient memory access during the computations as well as efficient communication of data, we want partitions at each level, if possible, to be stored in contiguous memory locations. For more exotic data structures, this means that packing and unpacking routines  may be required at the distributed level(s) in addition to the splitting scheme.
For (dense) matrices, defining an efficient storage scheme is relatively straightforward. Figure~\ref{fig:twoleveldata} shows an example of partitioning a matrix into two hierarchical levels. The matrix elements are ordered such that blocks (tiles) at each level are stored in contiguous memory blocks.

As described in Section~\ref{sec:versions}, accesses to shared data structures are managed through data handles. When a \ductteip data structure is initiated, handles are created for the data partitions at each level. In the same way as in the shared memory case, the handles track the versions of the corresponding data and host request queues for particular versions of the data. 



\subsection{Hierarchical task generation and task execution}
At level 0, the code is traversed by all participating computational nodes. Each task submission statement is inspected with respect to task ownership. The ownership of a task is decided based on the location of its output data. If the task is locally owned, it is generated and submitted to the level 0 run-time system.
 In the case of several outputs located in different computational nodes, a rule (e.g., first output) is implemented. Due to the hierarchical structure, the number of tasks submitted at the highest level is smaller than in a corresponding one-level approach, thereby reducing the overhead from the global inspection of the tasks in the initial task generation phase.

A task at level $\ell$ contains task submission statements for tasks at level $\ell+1$. When the level $\ell$ task is ready to run, the corresponding level $\ell+1$ tasks are generated and submitted to the level $\ell+1$ run-time system. 
Note that even though tasks at levels $\ell>0$ are submitted by different nodes, the order is globally preserved because these are submitted at the execution time of the parent task, which ensures that the dependencies at the parent level are respected and tasks are submitted in a consistent sequential order at each level. This is the mechanism that allows dependencies between any tasks scheduled by the same run-time, not only sibling tasks. (At the shared memory level, there is one \superglue run-time deployed in each computational node.)

The level $\ell$ and level $\ell+1$ tasks have a parent-child relationship. The task completion of the parent task depends on the completion of all of its children tasks. However, the run-time system handling the parent tasks can start the next ready parent task in the queue without waiting for completion of the previous one, as all ready tasks already have their dependencies fulfilled.

The tasks working on the last level data, contain the actual computational kernels. A benefit that follows from the hierarchical task submission is that the submission of lower level tasks is paced in relation to the speed of execution due to the synchronization with the parent tasks. This prevents the number of tasks in the work queues from growing in an uncontrolled way.
%
%


When tasks are placed according to the location of their output data as described above, the resulting load balance at the distributed level is static. If the data distribution scheme is not well chosen, this can result in severe performance degradation in the overall execution time of an algorithm. Distributed dynamic load balancing/scheduling~\cite{MGG17,moniruzzaman15,Gengbin05} is challenging as latency is much higher than for the shared memory case, and data movements both before and after the execution of a task must be considered. Another issue that becomes increasingly relevant as the scale of the distributed system increases, but which we will not discuss further here, is fault tolerant execution, see~\cite{Kurt14,lifflander14,BHBDD15,CHBD15,Ni16}.

\subsection{The data allocation scheme}
\ductteip creates a memory pool at each computational node at start-up time before the task submission begins. By using a memory pool, the overall time spent for memory allocation and deallocation is reduced as well as shifted out of the task execution phase. Within the memory pool, space is allocated for all the locally owned data and for data that will be received from other nodes during the task execution phase. The memory pool is dynamically extended when needed until the node runs out of memory. The pooling mechanism can be switched off for any individual data, and then allocation is implemented by the application code instead.
                          
Each data item in the memory pool resides in a contiguous memory block $M$ that contains a header part (of size $H$) and a content  part (of size $C$). The header part holds meta-data including the data handle with current version and memory address of the data, data size, and data ownership. A local task uses the content part directly in its computations. When a task at another node needs the data, the \ductteip run-time initiates a communication call for the header address and size $H +  C$. The data is received in a buffer within the memory pool of the remote node. The same memory area is used for further partitioning the data into smaller blocks, hence the received data can immediately be used by tasks at deeper levels (see Figure~\ref{fig:twoleveldata}).


When using data versioning in the distributed case, it is possible for two or more versions of the same data to exist at the same time. A new version of a data may arrive while an older one is still involved in a computation. Data that is received by a node gets a new place in the memory pool. Therefore, memory must also be released at run-time when a data version is no longer needed. When a newer version is accessed by a task, it means that all older versions are obsolete and the memory slots of these can be released to the pool.


Precisely how memory is allocated has implications on the NUMA effects during the execution. When the memory pool is allocated, the allocation is by default distributed evenly (interleaved in a round-robin mode by the NUMA controller) among the NUMA nodes (sockets/cores) of the computational node. If we instead allocate space for each data individually, the same NUMA allocation policy is applied per block. This is especially bad for the smaller data blocks used at the shared memory level. In the experiments, we can see a significant performance improvement when the memory pool is used, and finer partitions are derived in place as offsets from the parent data starting address compared with individual allocations. There are further possibilities to optimize the NUMA behavior that we have not explored so far. Using the \texttt{hwloc} library~\cite{BCMFGMTN10}, fine grained control over the association of memory with NUMA locations can be implemented.

\section{The distributed communication model}\label{sec:comm}
In a distributed task based programming model, it is expected that the necessary communication should be handled transparently by the run-time system. One drawback of this model can be that efficient implementations of global communication patterns like all-to-all or all-reduce are not immediately available. In the task based view, communication is pairwise and per task. A mixed programming model could be allowed, where user MPI code is interleaved with task-based regions. Here, we describe the constructs used in \ductteip for the point wise communication. 

\subsection{The listener concept}\label{sec:listener}
The object that we call \textit{listener} is used for managing communication at the distributed level(s). At task generation time, when data accesses are registered, the location of the data is also examined. In cases where a task requires a specific version of a remote data, a listener is generated by (or sent to) the data owner. This means that the remote node knows that the data will be needed later and can send it as soon as the required version becomes available. That is, the communication is initiated at the time the data should be sent, not earlier. In this way, we do not need to have a long list of outstanding communication requests. 

At execution time, when a data handle is upgraded to a new version, tasks that are waiting for that version are examined for execution. Additionally, the listeners waiting for that data version are processed, and the new data is sent to all nodes that have one or more listeners in the queue. Duplicate listeners can occur if several tasks hosted in the same remote node need the data, but then only one message is sent to avoid unnecessary communication traffic. 

Sending data corresponds to remote data accesses, and the data version needs to be updated in a consistent way. If the remote accesses are of the read type, the version can be upgraded immediately, while if the access involves modifying the data, the version cannot be changed until the modified data has been returned from the modifying node and received by the data owner. In the basic \ductteip model, the output data is always located in the node that is running the task, and then it is enough to count the number of listeners that are processed and upgrade the version accordingly. An example of the events associated with one listener is given below

\begin{itemize}
\itemsep=0pt
\item During \textit{task submission}, task $t_1$ owned by process $P_1$, that needs a remote data $d_2$ at process $P_2$ is encountered.
\item At $P_2$, a listener is created for the version $v$ of $d_2$ that is required by $t_1$.
\item During \textit{task execution}, version $v$ of $d_2$ becomes ready at $P_2$.
\item The listeners waiting for version $v$ are examined, and a message containing $d_2$ is sent from $P_2$ to $P_1$.
\item The version of $d_2$ is incremented, and the listener(s) are deallocated. 
\item The message with $d_2$ is received by $P_1$. Task $t_1$ waiting for version $v$ of $d_2$ is notified.
\item If all other data dependencies are fulfilled, task $t_1$ is ready to run. 
\end{itemize}


A benefit derived from having listeners is that it enables task and data migration/stealing. When a data is moved, listeners waiting for that data are moved along. Moving tasks can be done in different ways. In our 
ongoing work on dynamic load balancing, only ready tasks are moved, and then together with copies of their input data. It is assumed that a task that has been moved still returns the output data to the original data location.

\subsection{Asynchronous communication and threading models}

For the implementation of the communication model, the main objective is to overlap communication and computations. We also want to avoid stalling of the communication due to receive requests that are not fulfilled. An attractive solution would be to use multithreaded MPI, and let one thread manage send requests and one thread manage the receive state. Using more than one thread for MPI calls requires that the MPI library used is fully compliant with the MPI-2 specification of threading models. In particular \texttt{MPI\_THREAD\_MULTIPLE}, in which many threads can call MPI routines independently. 

If the MPI implementation is not fully compliant, the request for using \texttt{MPI\_THREAD\_MULTIPLE} may instead result in one of two other threading models. In the serialized threading model, \texttt{MPI\_THREAD\_SERIALIZED}, multiple threads can call MPI routines, but only one thread at a time is allowed to make progress. The others are blocked until the working thread completes. In the funnel threading model, \texttt{MPI\_THREAD\_FUNNEL}, instead only one thread is allowed to call MPI routines. 

When the current version of \ductteip was implemented, the latest versions of the most common MPI implementations were not fully compliant with respect to multi-threading. This means that in practice, only one thread could be used for communication. We could have chosen to split task management and communication over two threads, but since the task management and the communication are tightly connected, we decided to use only one thread per MPI process. To achieve overlapping of the task handling and the communication, all send and receive MPI calls are therefore in asynchronous non-blocking mode, using  MPI routines that return immediately. The state of the requests is also checked in non-blocking mode, which means that a periodic polling is needed.

The MPI process works in a loop until termination, where the send and receive states, the task states, and the listeners are checked. The MPI thread is co-located with the \superglue administration thread, that handles task submission, at the same core. In this way, computations and administration are interleaved at this core, while the threads at the other cores are left without interference. How well the asynchronous communication works also depends on how well the MPI threading model is implemented. In the experiments reported later, the communication is working well in this way. The listener based communication, where data is sent at the time it becomes available helps to reduce the risk for message flooding as requests are separated over time.










 
\subsection{Hybrid approaches versus pure MPI approaches}
We have chosen to use a hybrid MPI--Pthreads approach in \ductteipw. In this section, we describe the main reasons for this choice from a performance perspective.
Whether using a hybrid or a pure MPI approach, the parallel program runs on $p$  distributed computational nodes, each with $n$ cores. In the pure MPI case, $pn$ instances of the program run concurrently,  with $n$ instances within each node. In the hybrid MPI--Pthreads case instead $p$ instances of the program run on the computational nodes, and within every node the work is distributed over $n$ threads.

Using only MPI instances of a program to exploit the parallelism provided by the underlying hardware degrades the performance in comparison with a hybrid approach for the  following reasons: First, the memory allocated for each instance of an MPI process within a node is private to that process and can only be accessed by other processes via the main memory of the node, using techniques like inter process communication (IPC). Therefore, communicating even cached data between two processes running on the same CPU, and hence potentially sharing the same cache, requires a round-trip for the data to the main memory. 
%
Secondly, even if the communication within one node can be performed efficiently, the traffic between nodes is likely to increase as each of the processes within one node may need to communicate with each of the processes running at a remote node. In the hierarchical hybrid case, communication requests are instead aggregated at the node level.

%

Additionally, in the pure MPI case, the amount of memory that needs to be allocated by one node to receive messages becomes unnecessarily large. 
For example, if processes $P_1$ and $P_2$ running on node $i$ own data $A_1$ and $A_2$ respectively, and communicate these data between each other, then we need memory space for each of the two items at node $i$. When instead using a hybrid approach, no extra space is needed for communication within the node. This means that in a pure MPI mode, we run out of memory faster when increasing the node local work size. 


\section{Creating and tuning applications in the \ductteip framework}\label{sec:user}
Even if using a programming framework relieves the programmer from many of the more technical aspects of parallel programming, there are still things that the programmer can influence both regarding the algorithm and the program settings that affect the performance of the parallel software.

In this section, we start by discussing in general terms how to express an algorithm in a form that makes it suitable for task parallel execution. Then we take a closer look at the two algorithms that are used in the numerical experiments in this paper, Cholesky factorization and a solver for the shallow water equations. Finally, we discuss how a simple simulation tool can give insights concerning the expected performance. 

\subsection{Taskifying algorithms}
As explained in Section~\ref{sec:hier}, an algorithm in the \ductteip or \superglue frameworks is working on (hierarchically) partitioned data. The first step in preparing an algorithm for task-based parallelization is to decide how to partition the shared data. For matrices, blocks or slices are natural partitions. The algorithm then needs to be expressed as working on these data partitions. This could for example correspond to a transition from an element wise Cholesky decomposition algorithm to a block Cholesky factorization. 

A scalar control variable or a global vector that local results are gathered into can also be shared variables. Common for all shared data is that it is protected by a handle, which is used for managing accesses to the data.
 
Operations that can become tasks have a clearly defined interface to the shared data in terms of inputs and outputs, and do not modify any other shared data (no global variables). In the taskified code, such operations are replaced by task submission statements. The operations provide the computational kernel of the tasks. APIs for defining and submitting tasks are provided by the framework.

In the current implementation, the APIs for task submission and task definition with respect to the \superglue and \ductteip run-time systems have some small differences, and task definitions need to be provided for each participating run-time. Ideally, they should be the same in a general $L$-level framework. A unified interface that allows different run-times to cooperate transparently using a single application code is described, implemented and shown to perform well in~\cite{Zafari17}.

%

Since the \ductteip framework uses the same logic, concepts and abstracts as the \superglue framework does, any algorithm that can be implemented using  \superglue can similarly be implemented using \ductteipw. For some examples of algorithms implemented with \supergluew, see \cite{Tillenius15,HEGH14,TiLaLeFly15,BaEnWi16}. 

To learn more about the details of the framework implementations, as well as the application examples, visit the \superglue framework source repository at \url{https://github.com/tillenius/superglue/} and the \ductteip framework source repository at \url{https://github.com/afshin-zafari/DuctTeip}.


\subsection{Task parallel hierarchical Cholesky factorization}
The data in the Cholesky factorization algorithm is a dense symmetric matrix. Due to the symmetry, only the upper/lower part of the matrix is stored. The factorization is performed in place, such that the original matrix is overwritten by the factorization. The level 0 data is the whole matrix $A_0$. At level 1, the matrix is subdivided into $n\times n$ blocks. A natural formulation of an algorithm that works on the level 1 data is a block Cholesky factorization. The algorithm is given below in terms of calls to LAPACK subroutines.

\medskip

\begin{centering}
\begin{lstlisting}[language=C++,mathescape]
for(int i = 0; i < n; i++){
  for(int j = 0; j<i ;j++){
    syrk(A(i,j),A(i,i));          // $ A_{ii}=A_{ij}A_{ij}^T$
    for(int k = i+1; k < n; k++){
      gemm(A(k,j),A(i,j),A(k,i)); // $ A_{ki}=A_{kj}A_{ij}$
    }
  }
  potrf(A(i,i));                  // $ A_{ii}=LL^T$
  for(int j = i+1; j<n; j++){
    trsm(A(i,i),A(j,i));          // $ A_{ji}=A_{ii}^{-1}A_{ji}$
  }
}

\end{lstlisting}
\end{centering}

\medskip

The block Cholesky factorization contains four types of tasks \texttt{syrk}, \texttt{gemm}, \texttt{potrf}, and \texttt{trsm}. To continue to a hierarchical algorithm, each of the level 1 blocks are in turn subdivided into $m\times m$ blocks, and the algorithm of each task is expressed in terms of the level 2 data. The \texttt{potrf} operation is a Cholesky factorization, and is hence subdivided in the same way as the original algorithm, while the \texttt{gemm} operation is a matrix-matrix multiplication subdivided into \texttt{gemm} operations on the level 2 blocks. Figure~\ref{fig:choldag} shows an example of a level 0 task graph for a $4\times 4$ block Cholesky factorization and some of the level 1 subgraphs for a division into $3\times 3$ level 2 blocks.

The complexity of the dependencies between tasks and data in this factorization illuminates the flexibility of the framework in handling algorithms that are not straightforward to parallelize. In our implementation, we have not given priority to the \texttt{potrf} tasks which constitute the critical path. Priority can be included in the run-time scheduling approach to improve performance for algorithms where the critical path is known. 


\tikzset{
 nodest/.style={align=center, inner sep=0pt, text centered, font=\scriptsize,draw=gray,ellipse},
  potrfst/.style = {nodest,fill=fcpotrf},
  trsmst/.style = {nodest,fill=fctrsm},
  gemmst/.style = {nodest,fill=fcgemm},
  syrkst/.style = {nodest,fill=fcsyrk},
}
\begin{figure}[!htb]
\centering
\begin{tikzpicture}[>=latex',line join=bevel]
\node at (0,7) {\includegraphics[width=0.35\textwidth]{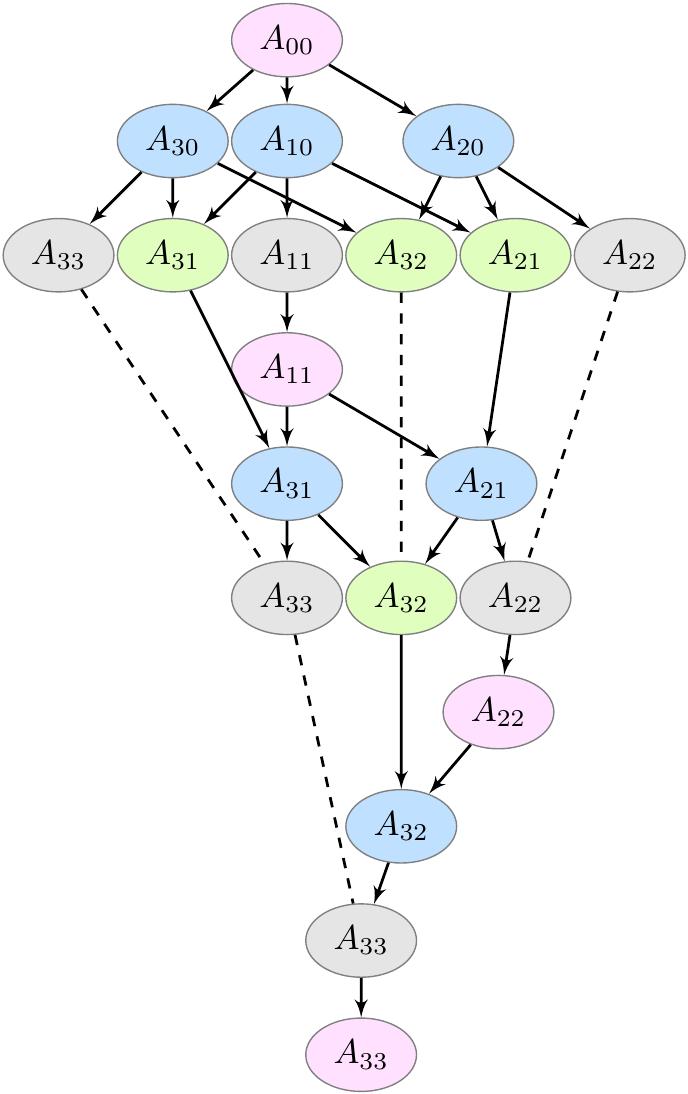}};
\node at (5,7.5) {\includegraphics[width=0.20\textwidth]{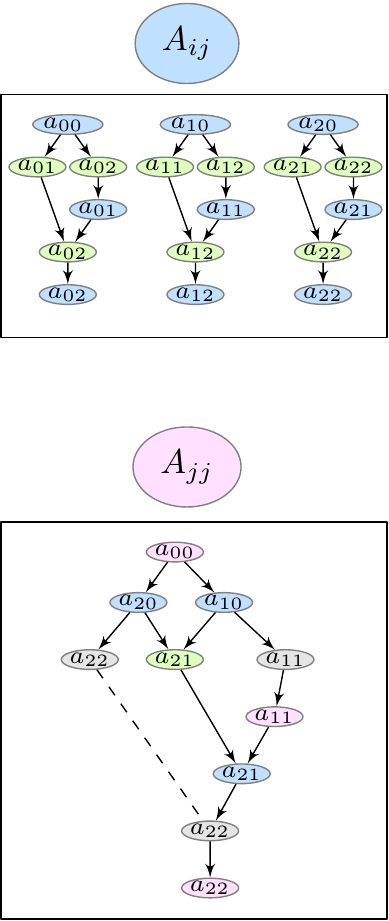}};
\node at (4,3.7) {\includegraphics[width=0.35\textwidth]{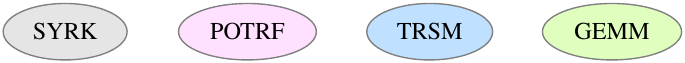}};
\end{tikzpicture}

\caption {To the left, a level 0 task graph of a $4\times4$ block Cholesky factorization of a matrix $A$ is shown. The labels of the nodes, $A_{ij}$, are the output data of the tasks. An arrow between tasks indicates a must execute before dependency, while the dashed lines indicate that the tasks can be reordered, but cannot execute concurrently. To the right, two examples of level 1 task graphs (a block TRSM and a block POTRF) for a $3\times3$ splitting of the level 1 data are shown.}
\label{fig:choldag}
\end{figure}


\subsection{A task parallel solver for the shallow water equations}
Dense linear algebra operations as the Cholesky factorization are common test cases for parallel implementation. They have certain advantages such as a large proportion of BLAS level 3 operations, which means that they are not so sensitive to the bandwidth limitations of modern hardware, they have regular data structures that allow for memory to be traversed linearly, and with a proper organization, the blocks are well suited for efficient communication.

Therefore, we wanted to include also a different type of application example. We have chosen a PDE solver for the shallow water equations on the surface of the Earth. The numerical method, based on radial basis function scattered node stencil approximations, is introduced in~\cite{FlyWri09}, and was parallelized using an MPI-enabled version of the SuperGlue framework (not using hierarchical tasks and data) in~\cite{TiLaLeFly15}. Here we instead use the hierarchical programming model of the DuctTeip framework.

The solver algorithm is data parallel in the sense that there is an explicit time stepping scheme, and each new value in the vector of unknowns can be computed based on data from the previous time step. However, the time stepping scheme is inherently sequential, which means that the potential parallelism has to be found in the spatial operator. The time-stepping scheme that is used is the fourth order Runge-Kutta scheme, and one time step has the following structure:
\begin{verbatim}
  f(F1,H);                  // F1 = f(H)
  add(H1, H , 0.5*dt, F1);  // H1 = H + 0.5*dt*F1
  f(F2,H1);                 // F2 = f(H1)
  add(H2, H , 0.5*dt, F2);  // H2 = H + 0.5*dt*F2
  f(F3,H2);                 // F3 = f(H2)
  add(H3, H ,     dt, F3);  // H3 = H +     dt*F3
  f(F4,H3);                 // F4 = f(H3)
  step(H,F1,F2,F3,F4);      // H  = H + dt/6*(F1+2*F2+2*F3+F4)
\end{verbatim}
The different sub steps within each time step also has a sequential dependency structure. The DuctTeip tasks for the time steps are submitted in two layers. There is one task for each time step, that in turn submits the tasks within that particular time step. If all time-steps were submitted in the beginning of the execution, there would be a problem with task overflow since the number of time steps in these simulations typically is very large. Therefore, the submission of time step tasks is controlled by a handle to a dummy data object that is owned by all processes. Each time step task has a modify access to that handle. A few time step tasks are submitted initially to fill the task pipelines, and after that, a new time step task is submitted whenever a previous time step task finishes.

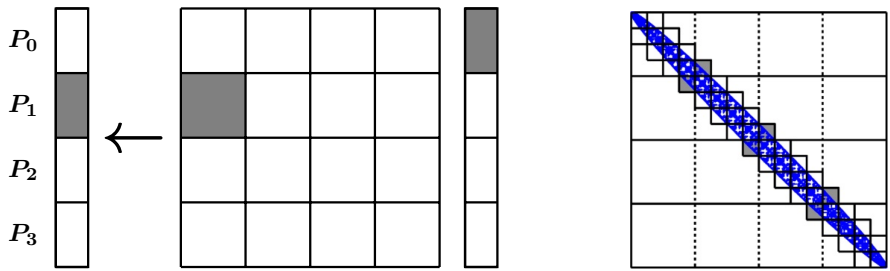
\begin{figure}[!htb]
\centering
\scalebox{.86}{
\boldmath%
\begin{tikzpicture}[line width=1pt]
\begin{scope}
\fill[white] (0,0) rectangle (.5,4);
\fill[gray] (0,2) rectangle (.5,2+1);
\draw (0,0) grid (.5,4);
\draw (0,0) rectangle (.5,4);
\end{scope}
\node at (1.7,1.95) [text width=2cm,font=\Huge] {$\leftarrow$};
\begin{scope}[xshift=55]
\fill[white] (0,0) rectangle (4,4);
\fill[gray] (0,2) rectangle (1,2+1);
\draw (0,0) grid (4,4);
\end{scope}
\begin{scope}[xshift=180]
\fill[gray] (0,3) rectangle (.5,4);
\draw (0,0) grid (.5,4);
\draw (0,0) rectangle (.5,4);
\end{scope}
\begin{scope}[xshift=0]
\node at (-0.5,3.5) {$P_0$};
\node at (-0.5,2.5) {$P_1$};
\node at (-0.5,1.5) {$P_2$};
\node at (-0.5,0.5) {$P_3$};
\end{scope}
\end{tikzpicture}}
\hspace*{15mm}
\includegraphics[width=0.27\textwidth]{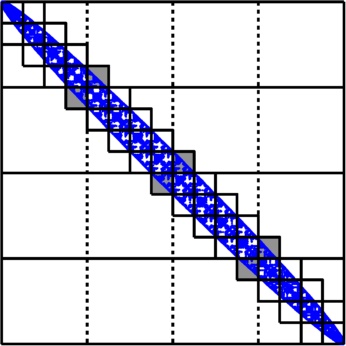}
\caption{Left: One block row of each data structure is assigned to each process $P_i$. The shaded blocks illustrate input and output in the block matrix multiplication. Right: An actual matrix structure together with a partitioning into level 1 blocks is shown. Tasks are only generated for blocks with non-zero elements. Tasks associated with the shaded blocks require communication between different processes.}  
\label{fig:blocking}
\end{figure}

The computationally heavy operations within a time step are the calls to the right hand side function \texttt{f(F,H)}, that evaluates the spatial part of the PDE operator. From the programming perspective, it consists of a number of sparse matrix-vector multiplications between differentiation matrices and instances of the vector of unknowns \texttt{H}. The data distribution of the matrix and vectors is shown in Figure~\ref{fig:blocking}.
%
The performance of the parallel implementation will depend on how well the sparse matrix-vector multiplication scales. The main challenges are the low computational density, which can lead to bandwidth contention and suboptimal scaling within a shared memory node, and the unstructured layout of the non-zero elements that leads to scattered memory accesses to the input vector, which is less efficient than linear accesses due to cache effects. For the sparse matrices, compressed row storage (CRS) is used. As shown in~\cite{TiLaLeFly15} the computational density can be increased by combining operations. There are four sparse matrices with the same structure that are multiplied with the four unknown vectors, and by storing the matrices and the unknowns together, the sixteen matrix-vector operations can be performed together. When increasing the number of computational nodes, we instead expect near optimal scaling since the amount of inter node communication is small, see Figure~\ref{fig:blocking}.

A critical aspect for this type of application is the synchronization points, because the number of tasks that can run in parallel is relatively small. The problem sizes that have been used for this particular application range from 6\,000 to 600\,000 data points in space. Since the hierarchical model employs synchronization between child tasks and their parent task, the potential mixing of tasks from different time steps, which can be beneficial for utilization, can be prevented if the number of tasks at the parent level are too few. 



\subsection{Simulation mode}
\label{sec:sim}

%

The actual performance of a task parallel application code will depend on how data is partitioned and distributed over the processes. This determines the amount of communication that has to be performed, how evenly the work is distributed over the processes, and how well the data sizes fit the memory hierarchy. 
In some cases, the properties of an algorithm are easy to understand theoretically and the best distribution can be derived analytically. In other cases, this may be less clear and some experiments are needed. To perform full simulations for many different parameter combinations to find the optimal ones is a waste of computational resources. A cheaper alternative is to use what is called \textit{simulation mode} in DuctTeip. The application code can be executed in simulation mode as it is, but the run-time replaces all computational kernels with an empty kernel and all messages with one byte messages. Statistics from the execution regarding numbers of tasks and messages are recorded to provide information to the user for deciding on partitioning and distribution schemes.



\begin{figure}[!htb]
\centering
\includegraphics[width=0.48\textwidth]{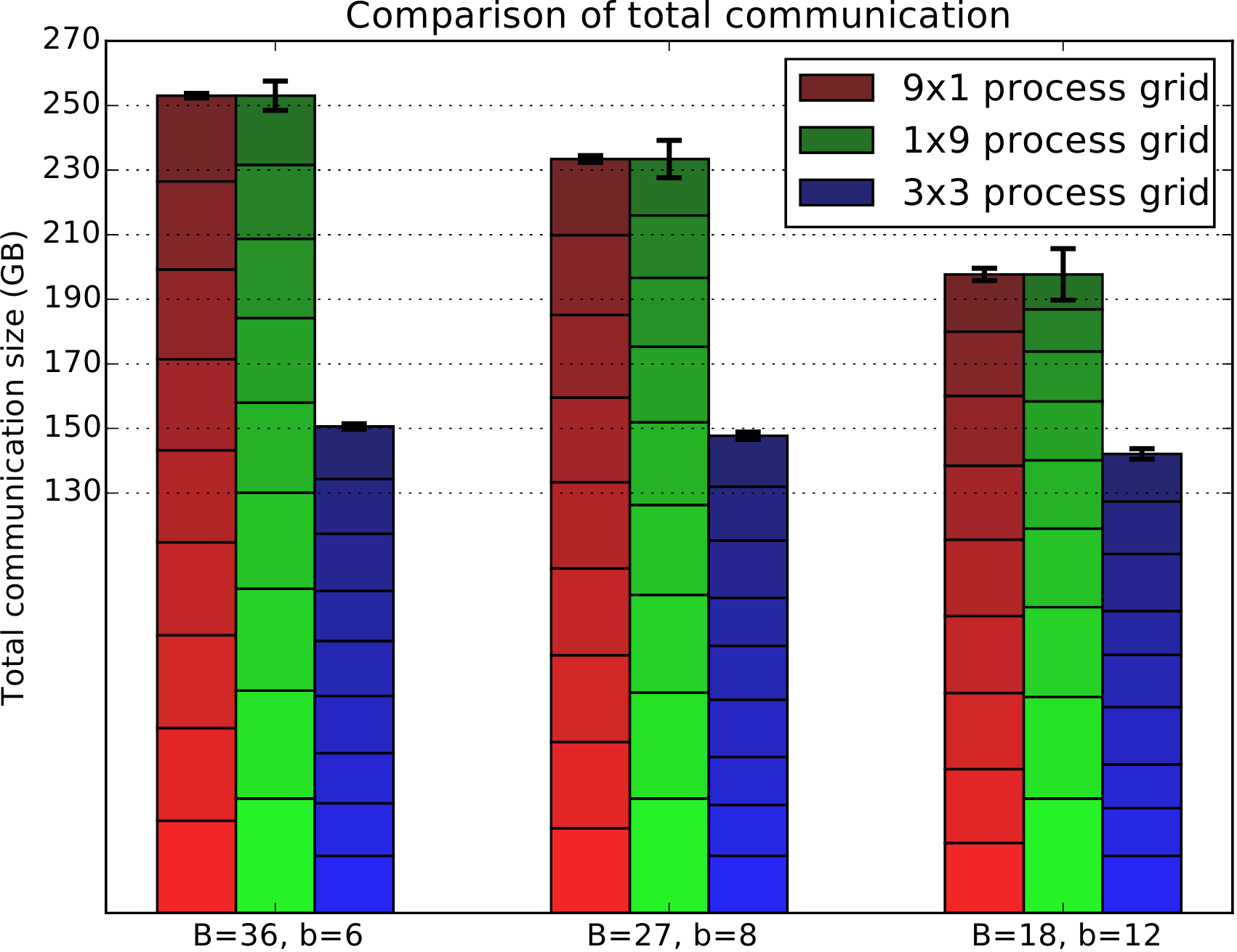}
\includegraphics[width=0.48\textwidth]{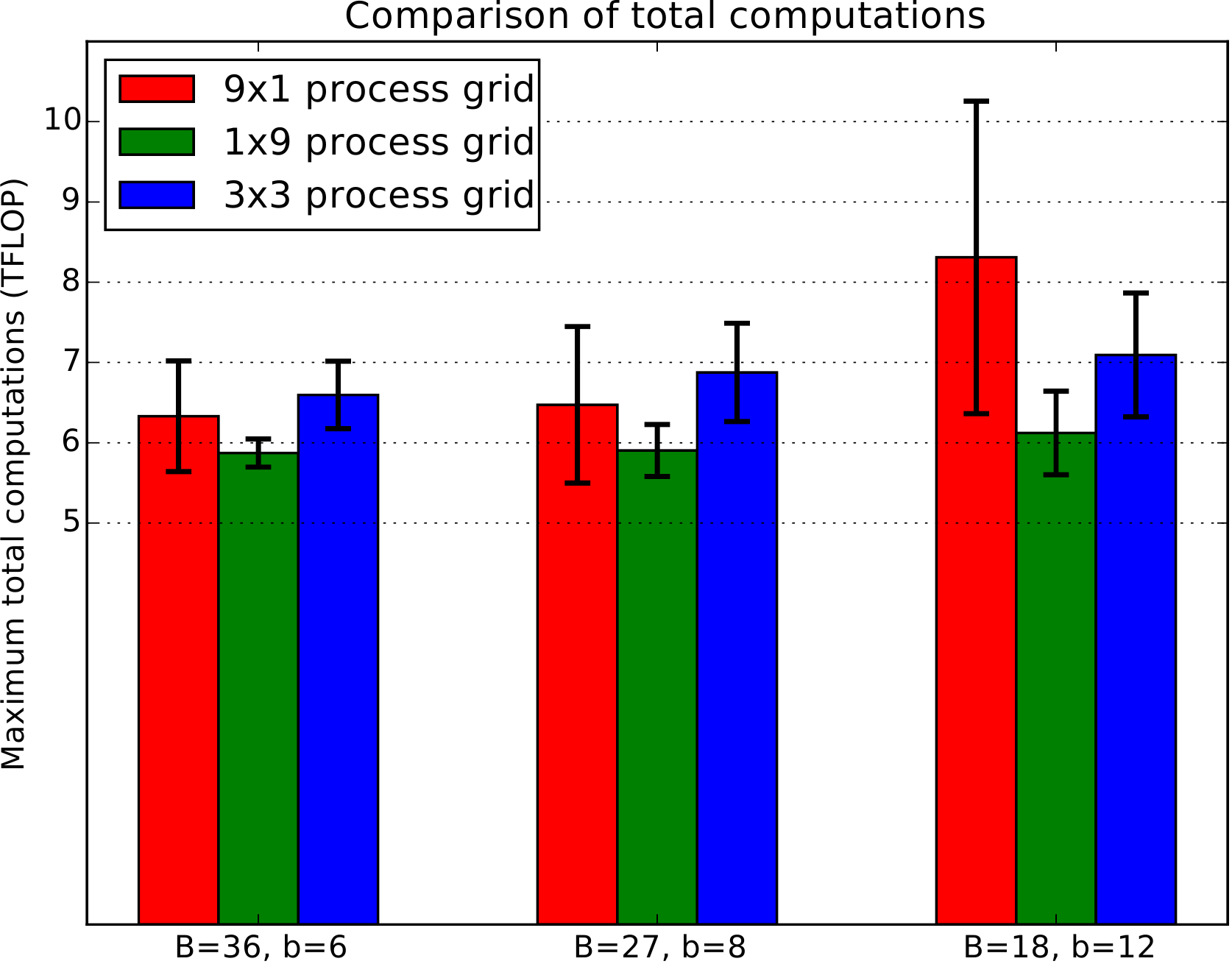}
\caption{Simulation statistics for Cholesky factorization of a matrix of size $N=B\times b\times 462 =99\,792$, where $B$ is the level 1 block size and $b$ is the level 2 block size. The simulation is executed on 9 nodes using three different process grids. The total communication is the sum of the number of GBytes communicated by each node, while the work is measured as the maximum number of level 2 tasks executed by one node multiplied by the typical work size, which is proportional to $n^3$ for the Cholesky factorization, where $n=462$ is the level 2 block size. The error bars show the variance over the nodes.}
\label{fig:sim_mode}
\end{figure}

To illustrate how this can work, we use the Cholesky algorithm as an example to see if we can recover the expected advice about how to best run it. Execution in simulation mode can not tell us what block size is appropriate for the caches. However, results for the level 2 block size can be obtained by running a small problem on one node. In all experiments shown here, this size is fixed to $462\times 462$ elements.   
Statistics for the total amount of communication and the maximum work on one node are given for three different process grids combined with three different blockings in Figure~\ref{fig:sim_mode}. 

Starting with the communication, we can observe that the total amount is the same for the $9\times 1$ and $1\times 9$ process grids, while it is significantly lower for the $3\times 3$ process grid. In all cases, the amount grows with the number of level 1 blocks, $B$, but only mildly for the $3\times 3$ process grid. Hence, from a communication perspective, a square process grid is desirable. Proceeding to the computational work performed by one node, the $1\times 9$ process grid has the smallest variance between nodes, and therefore also the lowest maximum work size. The $9\times 1$ process grid leads to smaller maximum work size than the $3\times 3$ process grid if $B$ is large enough, but suffers from significant load imbalance in the case $B=18$. In all cases, the work becomes more evenly distributed if the number of level 1 tasks $B$ is larger. 

%

The statistics for communication and computation point in different directions, but when comparing with actual run times, we have found that the communication size is the most informative measure. Having a large total communication size is likely to be detrimental to performance as the risk of tasks left waiting for remote data increases as well as the risk of congestion of messages. A square process grid is the factor that has the largest impact. Regarding the block sizes, having a large $B$ improves the load balance, but increases the amount of communication as well as the number of messages (another indicator that is not shown in the graphics).

\section{Experimental results}\label{sec:experiments}
The experiments presented in this section were performed during the time period when the main HPC cluster at the Uppsala Multidisciplinary Center for Advanced Computational Science (UPPMAX) was replaced. Therefore, the experiments where we compare the performance of different task parallel frameworks are performed at the Tintin cluster, while the experiments on the performance of the individual applications are performed on the Rackham cluster that replaced Tintin. We discuss how the properties of the different systems impact the results in connection with the specific experiments.



The Tintin cluster had 180 dual socket computational nodes with 64 GB/128 GB memory each. Each socket was equipped with an eight core AMD Opteron 6220 (Bulldozer) processor running at 3.0 GHz. The theoretical maximum performance for one Bulldozer processor is 192 GFlops and the theoretical memory bandwidth is 51.2 GB/s corresponding to 6.4 double precision loads per second. This means that $192/6.4 = 30$ floating point operations per load are needed to not suffer from bandwidth limitations.

The new Rackham cluster taken into production in April 2017 currently has 334 dual socket nodes with 128 GB/256 GB memory each. Each socket is equipped with a 10 core Intel Xeon E5 2630 v4 (Broadwell) processor running at 2.2 GHz. For a dual socket Broadwell node, the corresponding performance numbers are 704 GFlops and 136.6 GB/s leading to $704/(136.6/8)\approx 41$ floating point operations per double precision load. That is, to achieve a similar utilization as on Tintin, the computational intensity needs to be higher for Rackham.

\subsection{Weak scalability for the Cholesky factorization}
\label{sec:weakchol}
%
%

These experiments are run on the new Rackham cluster, and the application code is compiled with the Intel Parallel Studio XE 18.0.1 C++ compiler. When performing the parallel experiments, the problem size is increased with the number of nodes such that the amount of data that one node needs to communicate is fixed. The reason for fixing the communication size rather than the work size is that our experiment indicate that communication has a larger impact on the actual performance than the computations. The communication size is approximately proportional to $N^2/p$, where $N$ is the matrix size and $p$ is the number of processes (computational nodes). Hence, if we start with a matrix size $N_1$ for one node, then the matrix size for $p$ nodes becomes $N_p=\sqrt{p}N_1$. A square process grid is used in all cases, which is beneficial for reducing the total amount of communication, see Section~\ref{sec:sim}. The matrix is split into $B\times B$ level one blocks, and then each level 1 block is split into $b\times b$ level 2 blocks. 

\begin{figure}
\includegraphics[width=0.45\textwidth]{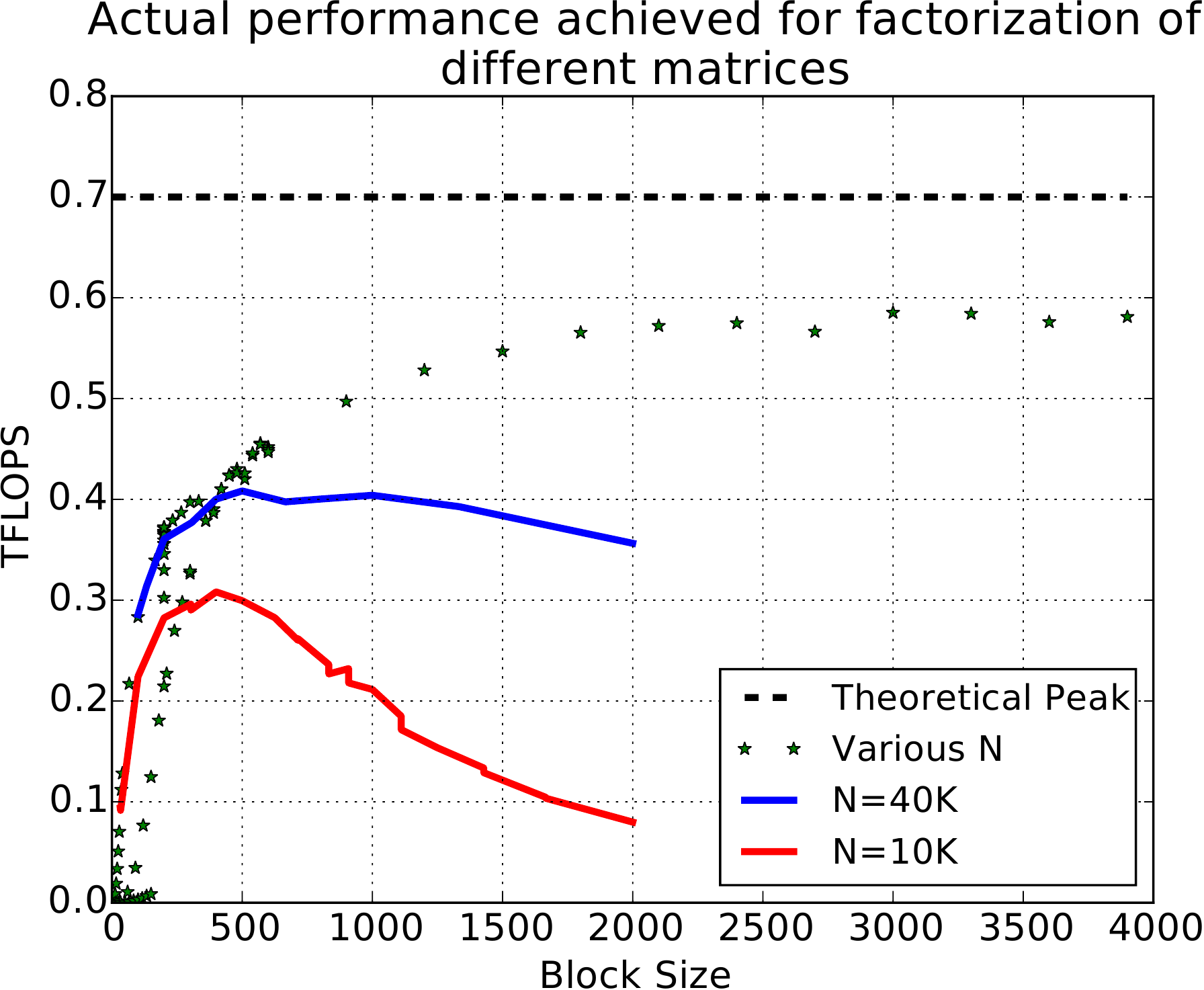}
\includegraphics[width=0.48\textwidth]{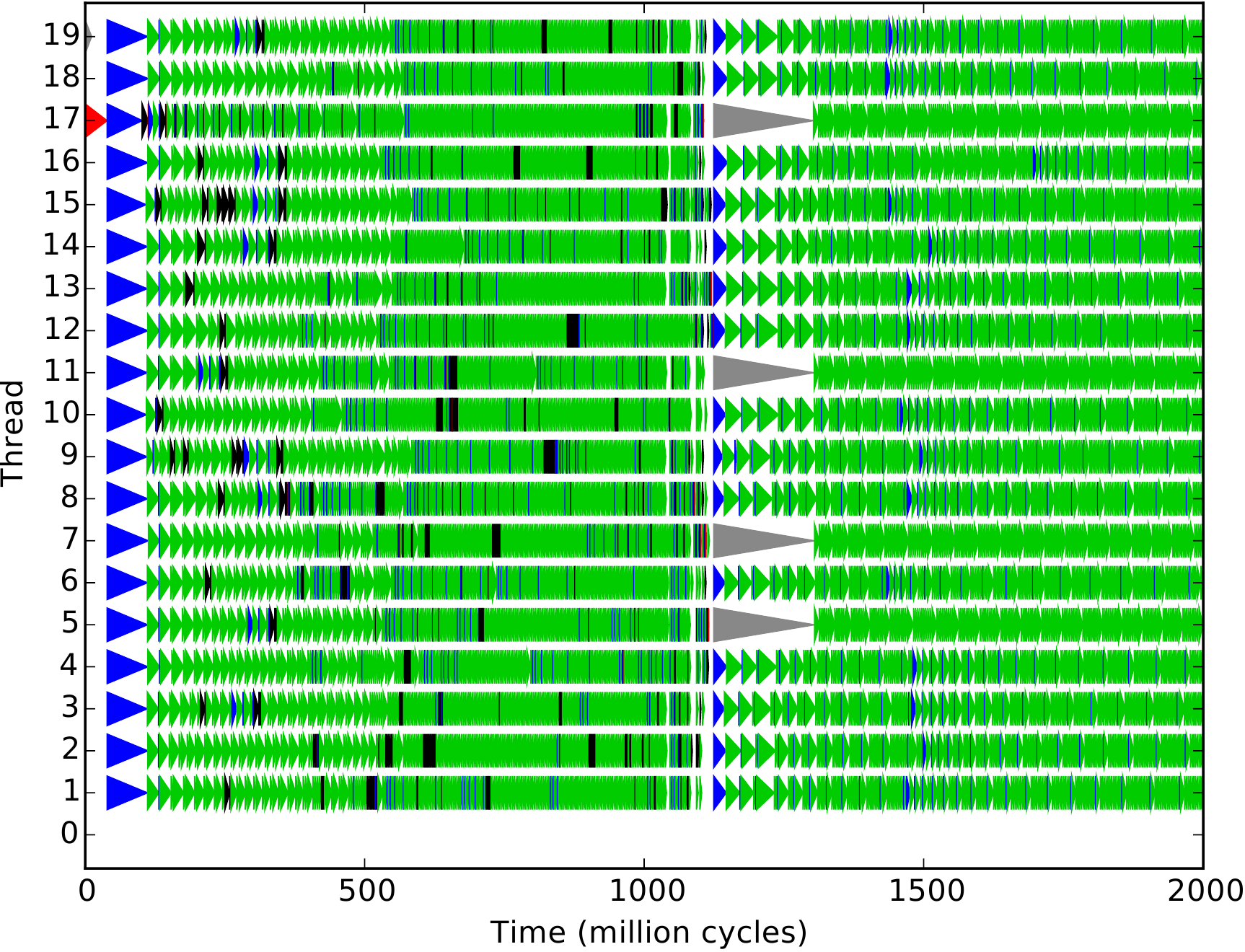}
\caption{Left: The computational speed as a function of the level 2 block size when the task parallel Cholesky application is executed on one computational node. Right: The beginning of the execution trace on one node for $N=40K$, $B=5$, and $b=30$. Each triangle represents a task, and the color indicates the type, with red for \texttt{potrf}, blue for \texttt{trsm}, green for \texttt{gemm}, and gray for DuctTeip parent tasks.}
\label{fig:cholperf}
\end{figure}

We start by investigating the performance on one node. The size $n$ of the level 2 blocks has a large impact on performance. It influences both how the blocks align with the cache sizes and the amount of computations per memory load. The left panel in Figure~\ref{fig:cholperf} shows how the computational speed varies with the block size. For the experiments marked with stars, the problem size is allowed to grow, while the number of blocks $B\times b=10\times 3$ is fixed. Then the computational speed increases with block size until it reaches a steady level. 
When we instead fix $N,$ we see that the performance follows the same curve for a while, but then performance goes down again. For $N=10\,000$, the performance starts going down for $n=500$ this corresponds to a product $B\times b=20$. The parallelism of the Cholesky task graph varies over the execution, see Figure~\ref{fig:choldag}, but for example, the first instance of the \texttt{trsm} operation has width $B\times b$ for a two-level graph. That is, the performance goes down when we no longer supply enough parallelism to keep the 20 threads busy. We are left with a trade-off situation, where the block size should be large enough for performance, but small enough for parallelism.
%

The best performance achieved on one node for block sizes $n\gtrsim 2000$ is about 85\% of the theoretical peak performance, which is computed by
 assuming that the software manages to use both vectorization (AVX instructions) and fused multiply-add (FMA) instructions fully. Then the Broadwell processors in Rackham can manage 16 floating point operations per cycle and core at 2.2 GHz, which gives a theoretical peak performance of 704 GFlops for one node.
%
Studying execution traces such as the right panel in Figure~\ref{fig:cholperf}, some of the causes for the distance from the theoretical peak can be understood. For the investigated problem sizes, thread 0 does not perform any computational work. Thread 0 handles task submission at the shared memory level, and if task execution is fast in relation to task submission (fine task granularity), it does not have time to do computational work. This stands for a 5\% performance loss in relation to peak. 
Another issue is the lack of parallelism in the beginning and the end of the Cholesky execution. The first panel factorization needs to be completed before more tasks become ready. When a relatively small $b$ is used, such as in the best performing results, there is not enough parallelism in this step to occupy more than a fraction of the threads. The rest are idle until the panel factorization completes. Similarly at the end of the execution the amount of parallelism decreases and there is some idle time among the threads. The first tasks also exhibit longer execution times than later tasks. This is likely due to memory effects, since data needs to be fetched from the main memory until the caches are warmed up. In problems where block sizes and block numbers are well chosen, the idle time for the actively working threads is small, around 0.5\%.

\begin{figure}
\includegraphics[width=0.45\textwidth]{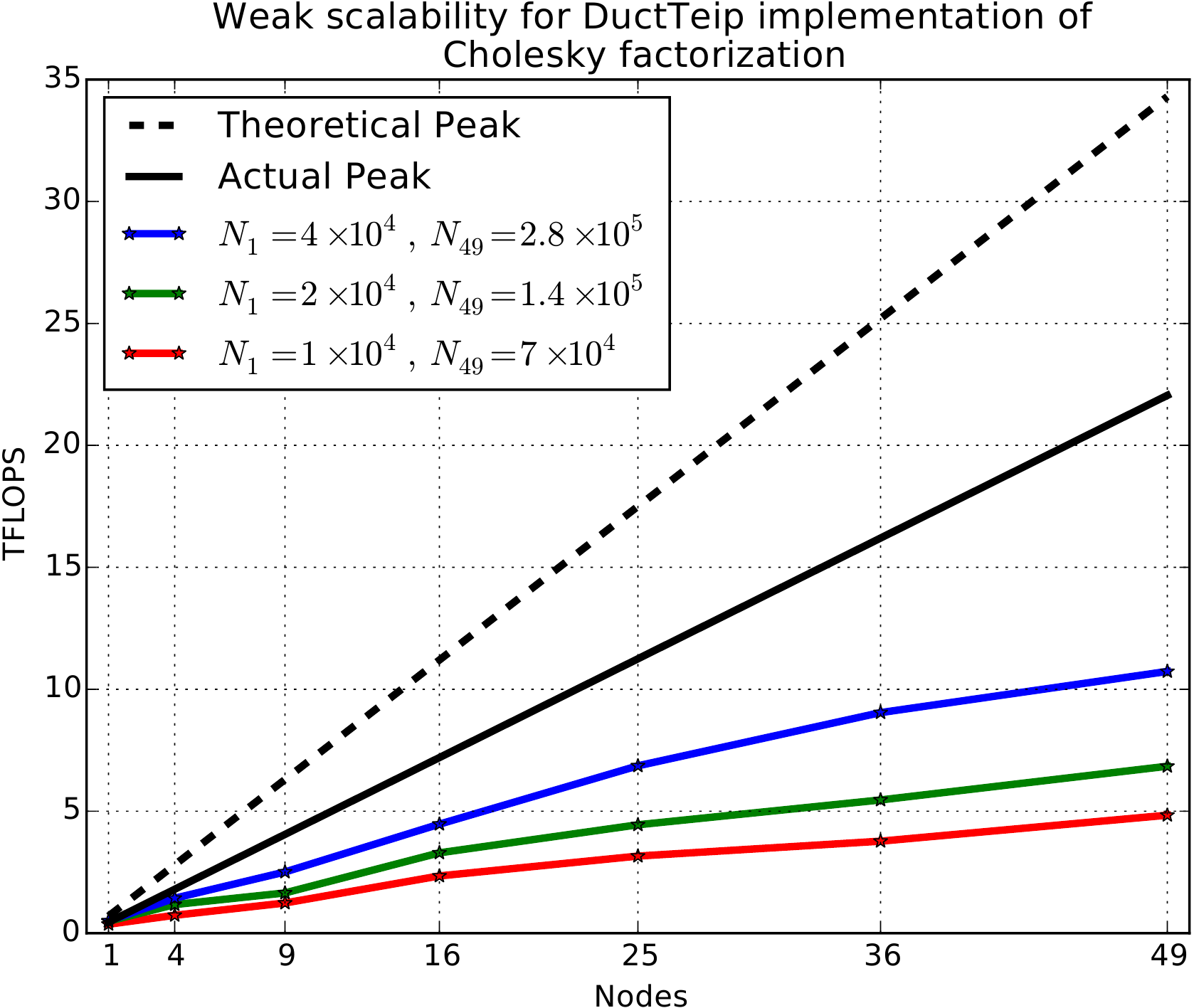}
\includegraphics[width=0.49\textwidth]{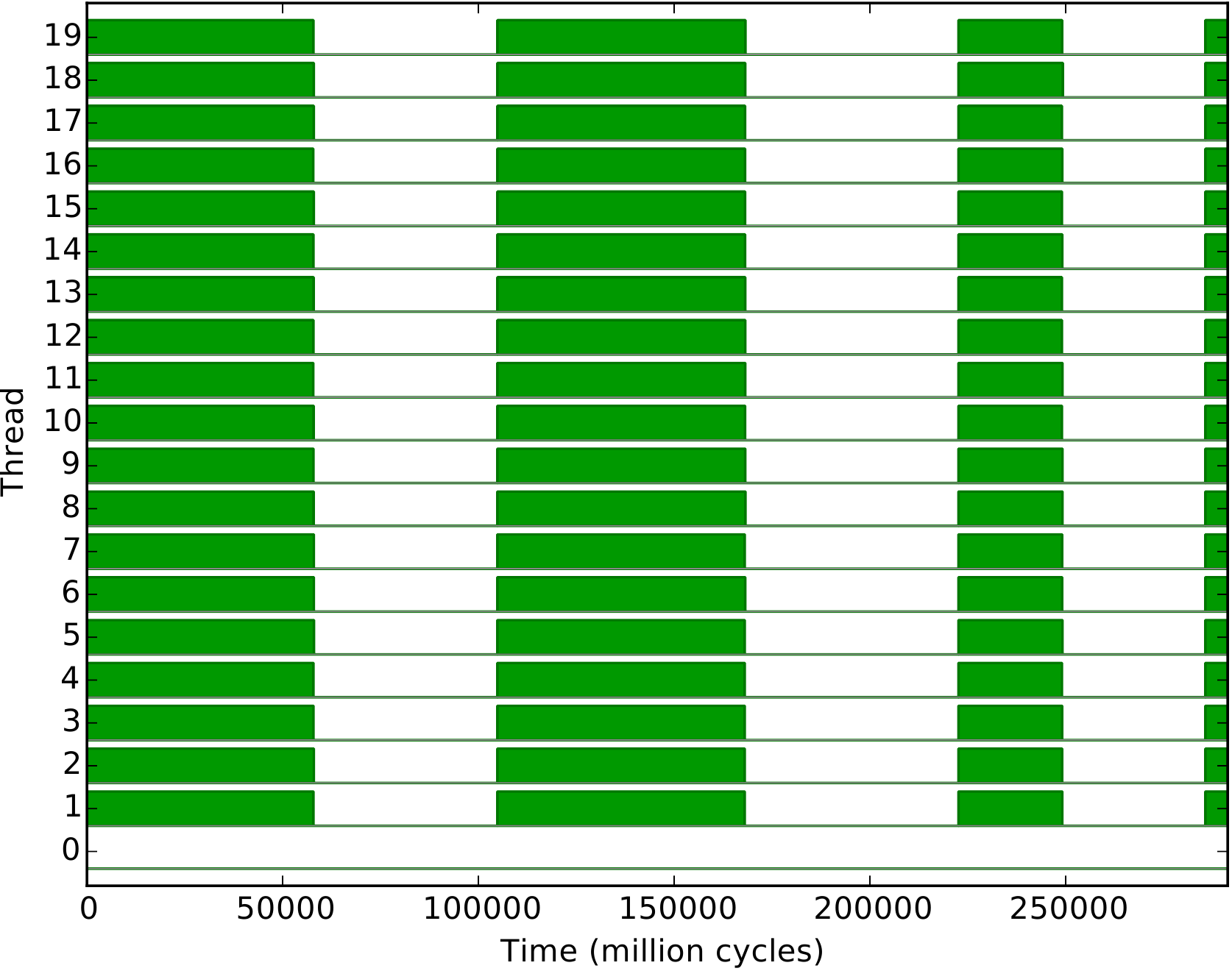}
\caption{Left: Weak scaling for three different initial matrix sizes. The theoretical peak corresponds to 704 GFlops/node while the actual peak is the speed on one node for the largest problem multiplied with the number of nodes. Right:
A simplified representation of the trace at node 0 when running the Cholesky application at 4 nodes. Busy time is indicated by filled blocks, while the idle time has just the 0 line. }
\label{fig:choltraces}
\end{figure}

For the multi-node weak scaling experiments, we have opted for having enough parallelism, which means that the block sizes are too small to give optimal performance. For the largest problem sequence, the typical block size is $n=300$, which means that we cannot expect more than around 0.4 TFlops/computational node. The parameter values used in the experiments are given in Table~\ref{tab:cholB}. 
\begin{table}[!htb]
\centering
\caption{The blocking parameters used in the Cholesky weak scaling experiments.}
\label{tab:cholB}
\begin{tabular}{|c|rrrrrrr|}\hline
$p$ & 1 & 4 & 9 & 16 & 25 & 36 & 49\\\hline
$B$ & 5 & 8 &12 & 16 & 20 & 18 & 28\\
$b$ & 30 & 30 & 30& 30& 30& 20& 30\\\hline
\end{tabular}
\end{table}
The right panel of Figure~\ref{fig:choltraces} shows the weak scaling results for the multinode experiments for three different initial matrix sizes. We can observe that the absolute performance improves with matrix size. The efficiency $E_p$ of the scaling, computed as $S_p$, the actual speed for $p$ processes, divided by the theoretical speed for linear scaling 
\[E_p = \frac{S_p}{pS_1},\]
is given in Table~\ref{tab:eff}.
For the largest matrix size we achieve about 50\% of the expected performance when increasing $p$, while the smaller problem sizes have a reduction also in efficiency.
 The traces again reveal the reason for the results. In the right panel of Figure~\ref{fig:choltraces}, we see the trace for the smallest matrix size running on four nodes. There are a number of gaps in the execution. We believe that these are connected with the inter-node communication, and that the computational work is not enough to hide the latency of the communication operations. There could potentially be some improvement through introducing priority for the critical path to ensure that data is communicated as early as possible. 

A positive aspect of the results is that the code scales up to 49 nodes for the tested problem sizes. There is some reduction in efficiency with the number of nodes, but the main loss is between running on one and multiple nodes.


\begin{table}[!htb]
\centering
\caption{The efficiency in the weak scaling experiments for the Cholesky application for different initial matrix sizes $N_1$. The speed on one node, $S_1$, is measured in TFlops.}\label{tab:eff}
\begin{tabular}{|r|r|rrrrrr|}\hline
$N_1$ & $S_1$ & $E_4$ & $E_9$ & $E_{16}$ & $E_{25}$ & $E_{36}$ & $E_{49}$\\\hline
10\,000 & 0.33 & 0.50 & 0.38 & 0.40 & 0.34 & 0.29 & 0.27\\
20\,000 & 0.44 & 0.60 & 0.38 & 0.42 & 0.36 & 0.31 & 0.29\\
40\,000 & 0.46 & 0.71 & 0.55 & 0.55 & 0.54 & 0.50 & 0.43\\\hline
\end{tabular}
\end{table}




\subsection{Strong scalability for the shallow water application}
For the shallow water solver, we look at the performance for fixed problem sizes. These problems are benchmarks for global climate simulations and are typically run for a certain resolution of the Earth. The largest matrix size we are using is $N=612\,416$, which is not very large for a sparse matrix. The main reason that the problems still take a long time to run is the large number of time steps that are needed. For example, for a simulation over 6 days with a time step of 30 s, 17\,280 time steps are needed. In the numerical experiments, we perform 20 time steps, which is long enough to capture the long term behavior of the solver. The block sizes used in the experiments are given in Table~\ref{tab:sw_block}.
\begin{table}[!htb]
\centering
\caption{The blocking parameters used for the shallow water experiments.}
\label{tab:sw_block}
\begin{tabular}{|c|rrrrrrrrrr|}\hline
$p$   &  1 &  2 &    3 &    4 &    5 &    6 &    7 & 8 & 9& 10\\ \hline
$B$   & 5 & 6 & 15 & 40 & 50 & 60 & 35 & 32 & 36 & 50\\ 
$b$  & 18 & 18 & 5 & 5 & 5 & 5 & 5 & 8 & 5 & 4\\ \hline
\end{tabular}
\end{table}

\begin{figure}[!htb]
\centering
\includegraphics[width=0.48\textwidth]{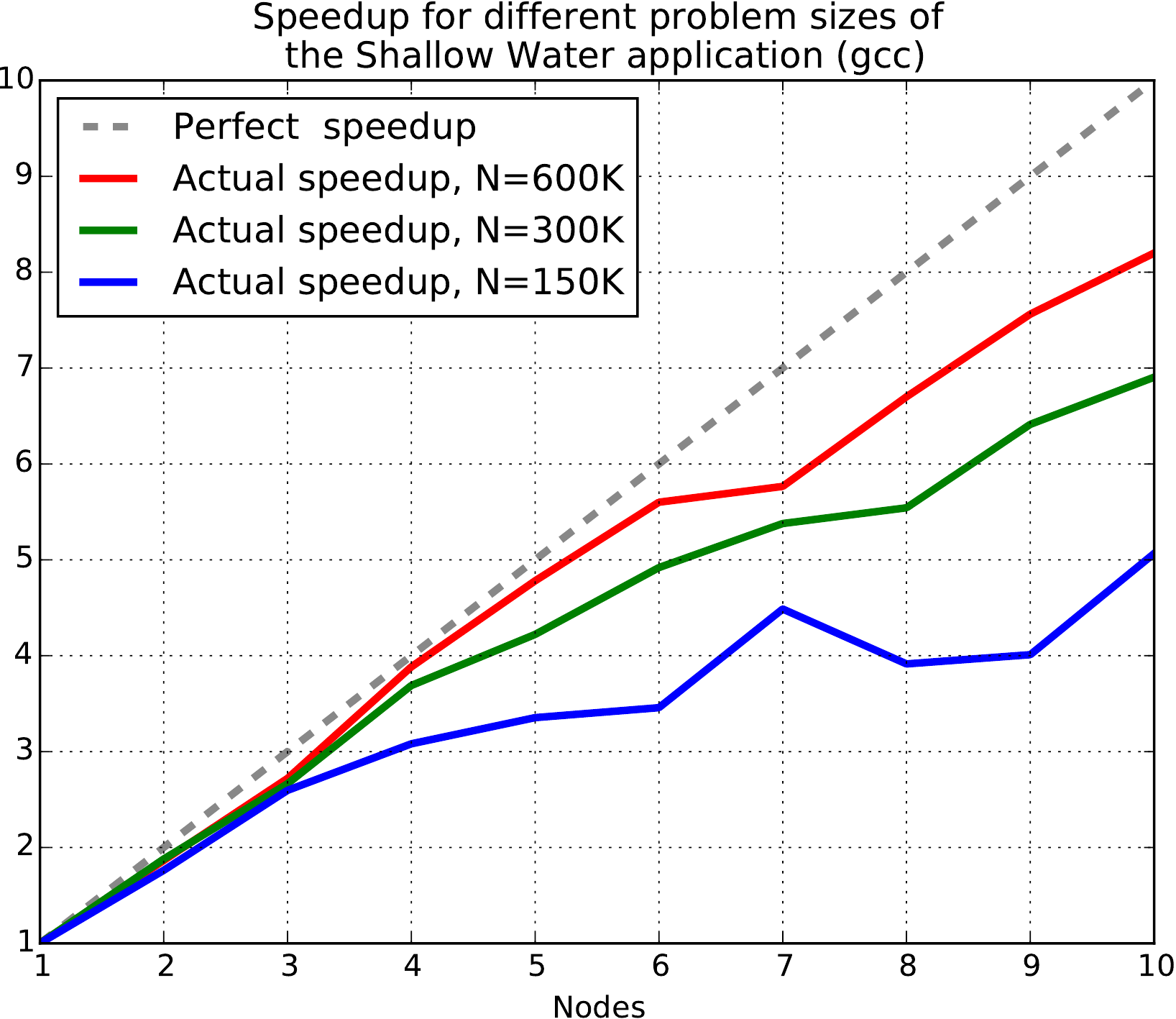}
\includegraphics[width=0.48\textwidth]{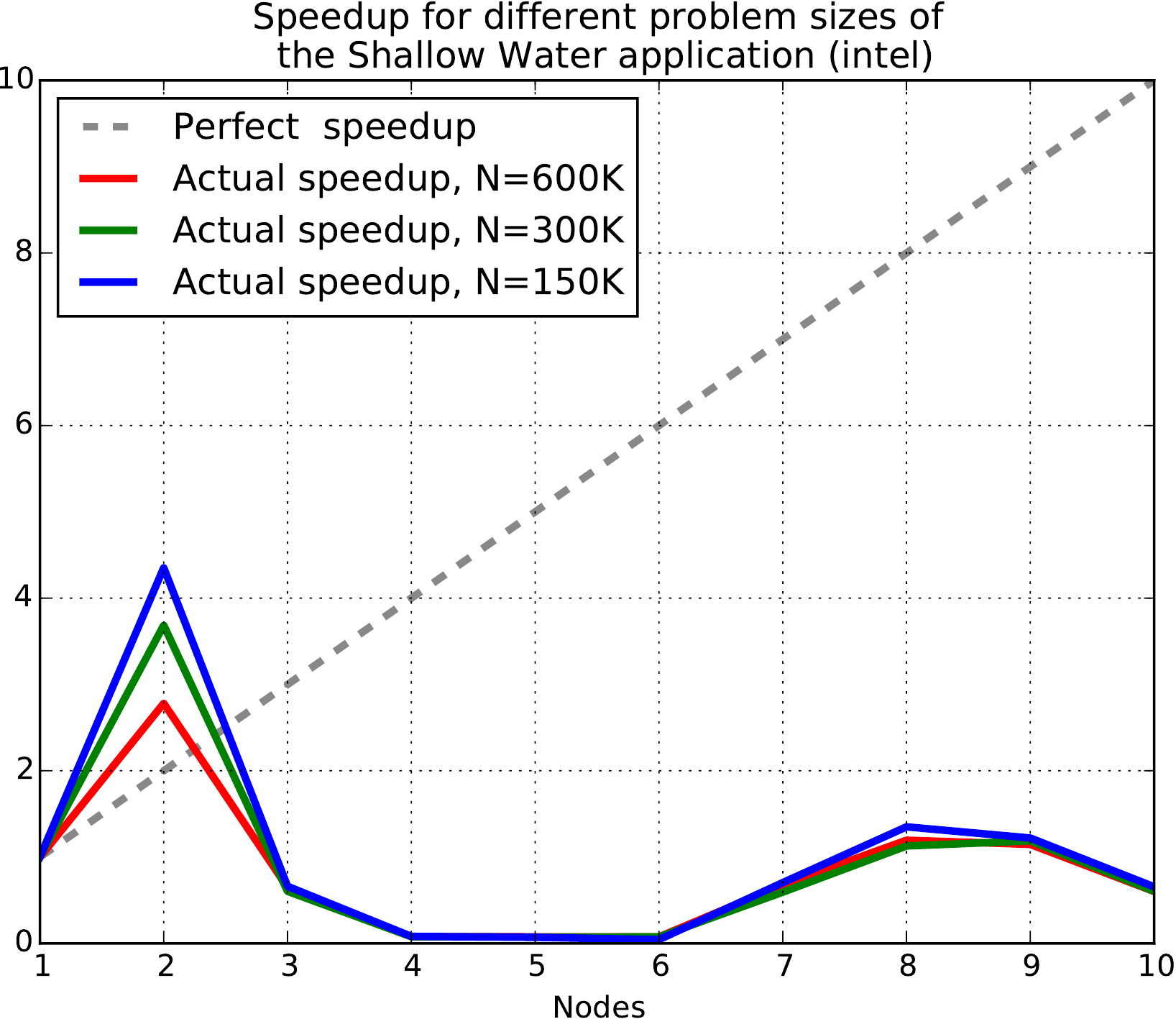}
\caption{Speedup for the shallow water application using the gcc (left) and Intel compilers (right). }
\label{fig:sw_speedup}
\end{figure}

First we look at the speedup $S_p=T_1/T_p$, where $T_p$ is the execution time on $p$ nodes. The left part of Figure~\ref{fig:sw_speedup} shows the results when compiling with gcc 7.2.0 and linking with OpenMPI. The speedup is close to linear for small numbers of nodes, and gradually moves away from linear when the node number increases. The speedup improves with problem size. However, the picture is different in the right part of Figure~\ref{fig:sw_speedup}, where the Intel C++ compiler has been used together with Intel MPI. Execution is unexpectedly fast at two nodes, but in all other cases, it is slower than or similar to execution on one node. Furthermore, the speedup at two nodes is highest for the smallest problem size.

To understand the differences between the two versions, we continue with comparing the absolute execution times for the largest problem size. Table~\ref{tab:sw} shows that the smallest run time is actually the same for both compilers, but it is achieved at 10 nodes for gcc, and for two nodes by the Intel compiler. Also, the run time on one node is three times smaller for the Intel compiler. One expected effect of this is that it will be harder for the code compiled with the Intel compiler to scale since the perceived work size is smaller.
\begin{table}[!htb]
\centering
\caption{Run times in seconds for the shallow water application with $N=612\,416$ when the code is compiled with the gcc (g) and Intel (I) compiler, respectively.}
\label{tab:sw}
\begin{tabular}{|c|rrrrrrrrrr|}\hline
$p$   &  1 &  2 &    3 &    4 &    5 &    6 &    7 & 8 & 9& 10\\ \hline
$T_p^\mathrm{(g)}$   & 60.4 & 32.4 & 22.4 & 15.6 & 12.6 & 10.8 & 10.5 & 9.0 & 8.0 & \textbf{7.4}\\ 
$T_p^\mathrm{(I)}$  & 20.3 & \textbf{7.3} & 31.5 & 257.9 & 269.4 & 273.7 & 30.8 &34.1 &35.5 & 33.8\\ \hline
\end{tabular}
\end{table}

\begin{figure}[!htb]
\centering
\includegraphics[width=0.48\textwidth]{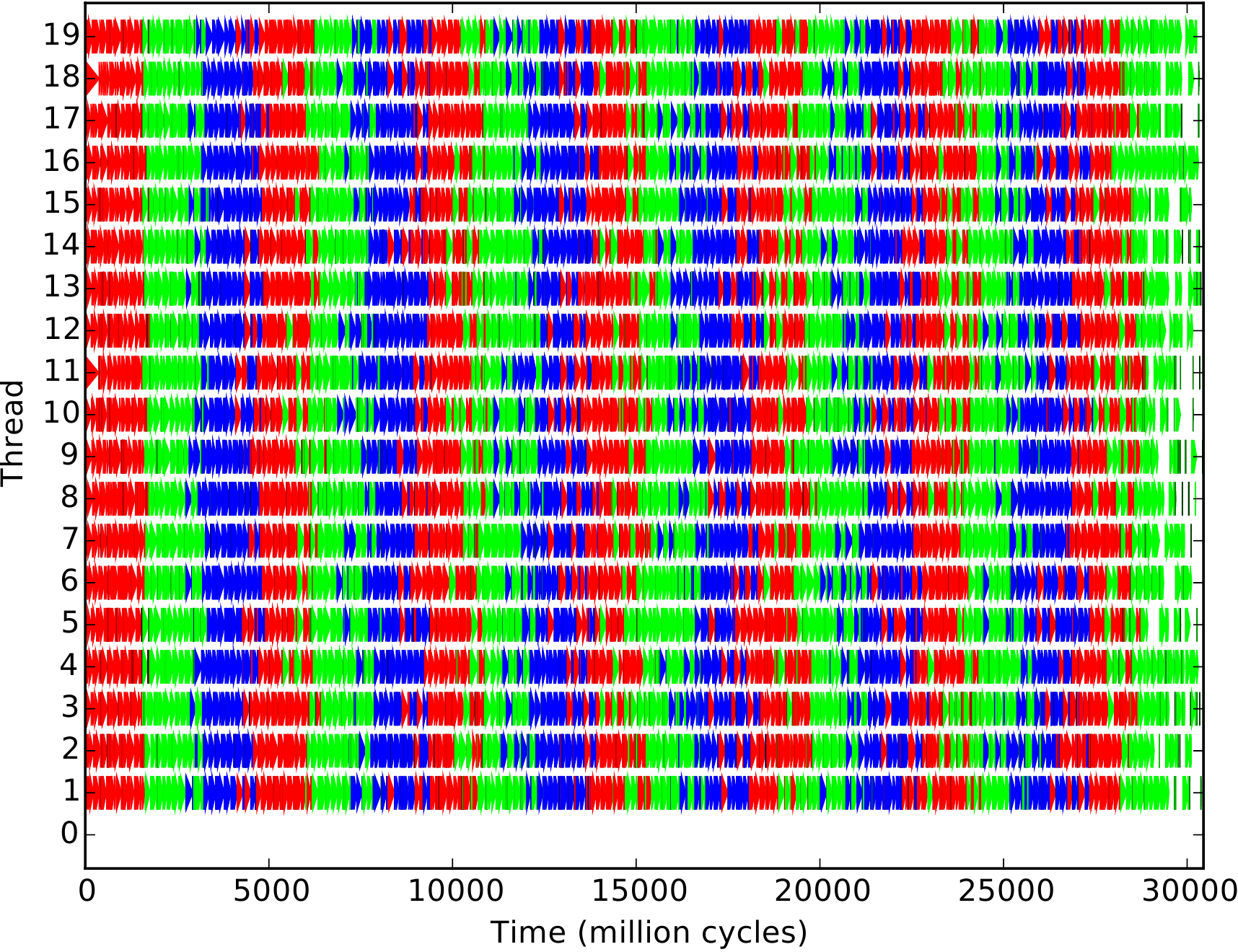}
\includegraphics[width=0.48\textwidth]{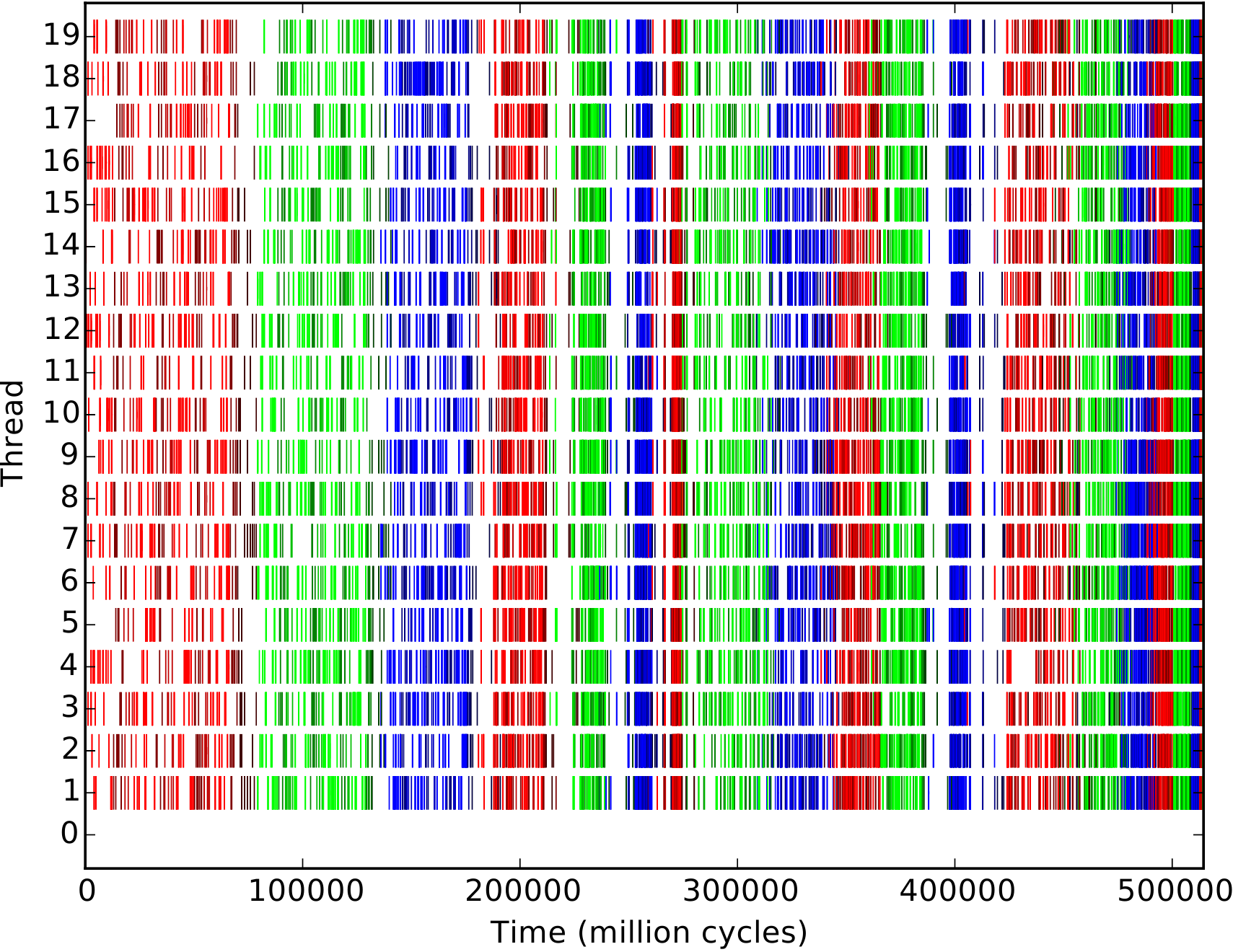}
\caption{The execution traces for node zero for the code compiled with gcc (left) and for the code compiled with Intel (right) running on four nodes. The matrix and vectors are here sliced into $B=40$ and $b=4$ and $b=5$ blocks respectively.}
\label{fig:sw_trace}
\end{figure}

Finally, we take a look at the execution traces. In the gcc case, all the traces have the same appearance, but with a different time scale. An example is shown in the left part of Figure~\ref{fig:sw_trace}. Thread 0 is occupied with task submission during the whole execution. The computations are performed on the remaining 19 threads. The different time steps are easy to identify, and the desired mixing that allows the work to be balanced over the threads is visible. This works as long as the number of tasks in each time step at both the DuctTeip and SuperGlue levels are large enough to provide parallelism.

In the right part of Figure~\ref{fig:sw_trace}, we show a trace for the code compiled with the Intel compiler. It is clear that there is significant idle time between the tasks. There is also a clear variation in how long time one time step takes. 
To further quantify what we see in the traces, we compare the time spent in tasks for the two versions running on four nodes. The tasks are on average computed 13 times faster in the Intel version than in the gcc version, in spite of this, the overall execution time for the gcc version is more than 16 times faster.
%

The drop in performance experienced with the Intel compiler can be related to observations made for SuperGlue regarding small task sizes, see~\cite[Fig.~6]{Tillenius15}. When the task size is too small, and the number of threads is comparatively large, the work stealing algorithm used for load balancing results in contention between the threads, preventing productive work.









\subsection{Comparison between hybrid and pure MPI parallel implementations}
							


In this section, we try to quantify the communication disadvantage of having a pure MPI implementation instead of the hybrid implementation using Pthreads within the shared memory nodes with an example. We run the Cholesky factorization code on two computational nodes with a  $6\times 6$ process grid in the pure MPI case (a square process grid is optimal), and a $2\times1$ process grid for the hybrid implementation.

Table~\ref{tab:hybrid-pure} shows the effect on the number of messages and the resulting execution times for the two cases. The total number of messages increases significantly with the increase in the number of processes for a pure MPI implementation. Furtermore, in the Cholesky case, the number of messages between the nodes also increases. The number of messages per process is however relatively stable. A negative effect that will become even worse when the number of nodes is increased is the length of the message queues. Here, the increase in queue length is 3.5, which has an effect on the time it will take to process messages and hence the communication overhead. The execution time in the pure MPI case is increased by 50\%. As we are actually using only 36 cores compared with 40 for the hybrid case, we should expect an 11\% increase, but not as much as 50\%.  
\begin{table}[!htb]
\centering
\caption{Execution statistics for a Cholesky factorization of a matrix of size $N=90\,000$ comparing the hybrid DuctTeip implementation and a pure MPI version.}\label{tab:hybrid-pure}
\begin{tabular}{|l|rrr|}\hline
                & Hybrid & Pure MPI & Ratio MPI/Hybrid\\\hline
Total messages & 16 & 455 & 28.4\\
Max messages/process & 12 & 20 & 1.7\\
Max pending messages & 2 & 7 & 3.5\\
Execution time (s) & 18 & 27 & 1.5\\\hline
\end{tabular}
\end{table}


\subsection{Benchmarking against other frameworks}
When comparing DuctTeip against other frameworks, we use the Cholesky factorization as a benchmark because this is a standard test case, which has been implemented also by the other framework developers. We compare against PaRSEC, Cluster OmpSs and StarPU, but we also include ScaLAPACK to represent an application specific implementation with an optimized static parallel schedule for the work. To be compatible with ScaLAPCK that uses a square process grid, we only use square numbers of nodes in the experiments. As mentioned in the introduction to the experimental section, these experiments are performed on the older cluster Tintin.

Compilers and libraries have been selected to provide the best possible performance for the target hardware at the time when the experiments were made. Since the software is continuously upgraded, improved versions may be available. Below, we specify the choices for the frameworks and other components that were used.
\begin{description}
\itemsep=0pt
\item[BLAS:] The ACML 5.3.1 implementation of BLAS is used, except for StarPU, which uses the ATLAS library 3.10.2, in both cases the libraries are optimized for the target system.
\item[MPI:] All applications are linked with OpenMPI 1.6.5.
\item[hwloc:] Version 1.9.1  is used by PaRSEC and StarPU and built by gcc 4.4.7.
\item[DuctTeip:] Compiled with the Intel C/C++ 13.1 compiler.
\item[Cluster OmpSS:] The customized compilers Mercury 1.99.7 and Nanos+ 0.9a from the Barcelona Supercomputing Center are built on top of gcc 4.4.7. The Cholesky application is then compiled and built using these.
\item[ScaLAPACK:] Version 2.0.2 is compiled and built with the PGI 13.8 FORTRAN compiler and optimized for the target system. The Cholesky factorization is performed by calling the DPOTRF subroutine.
\item[StarPU:] Version 1.1.5 is built using gcc 4.8 and calibrated for the Cholesky factorization program using the dmdar (deque model data aware ready) task scheduling policy.
\item[PaRSEC:] DPLASMA 1.2.1 is compiled and built by the Intel C/C++ 13.1 compiler.
\end{description}

%
%

The most frequent operation in the block Cholesky factorization is a GEMM operation (matrix--matrix multiplication). To have a reference to what could be considered the maximal achievable performance we include what we call the GEMM peak in our comparisons. The maximum throughput of the \texttt{dgemm} BLAS subroutine on one node (16 cores) that is achieved on the Tintin cluster is 162 GFlops for square matrices of size $N=24\,000$. The GEMM peak is then computed as the one-node-result scaled by the number of nodes.





\subsubsection{Strong scaling experiments}\label{sec:strongscaling}
\begin{figure}[h!]
\includegraphics[width=0.5\textwidth]{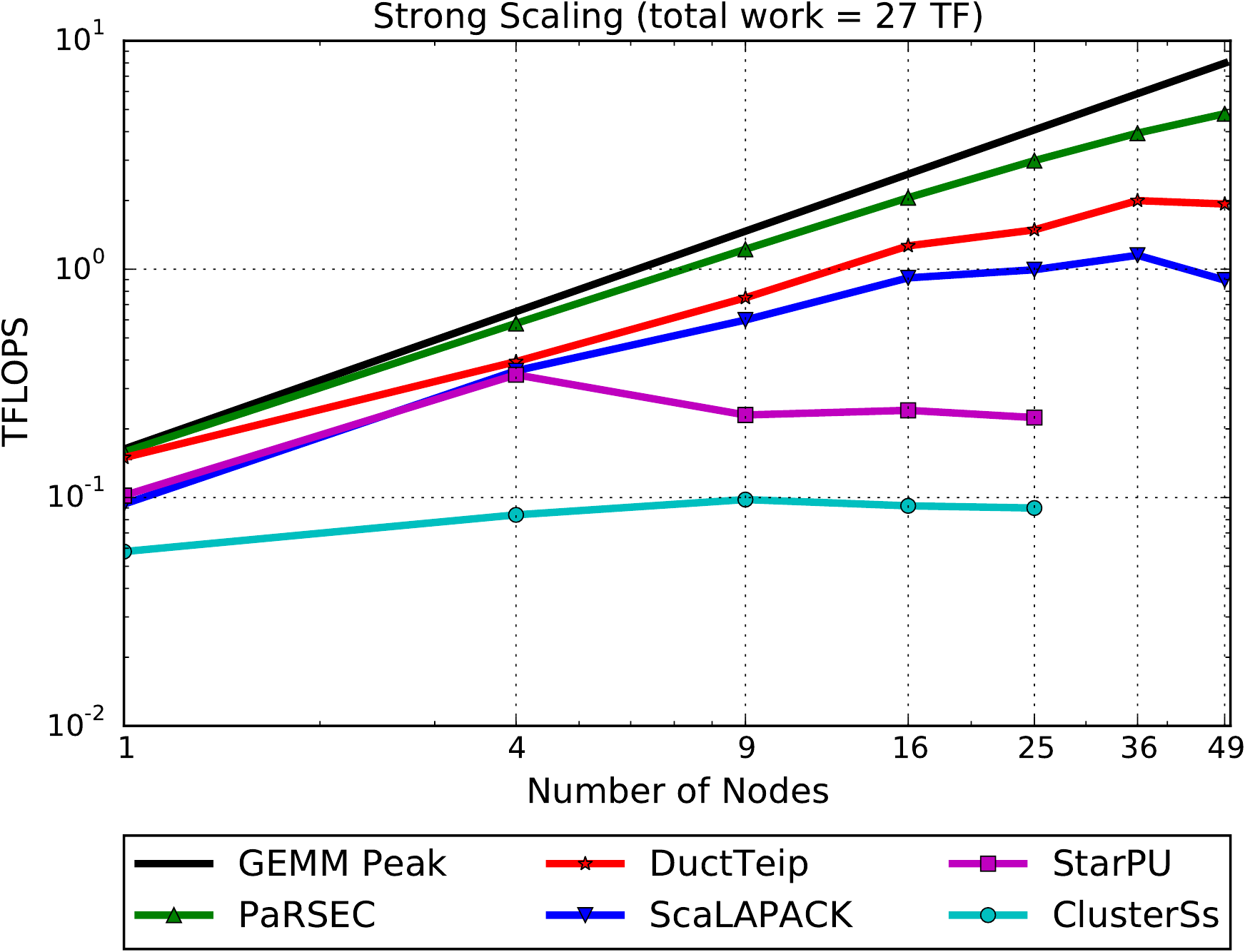}
\includegraphics[width=0.5\textwidth]{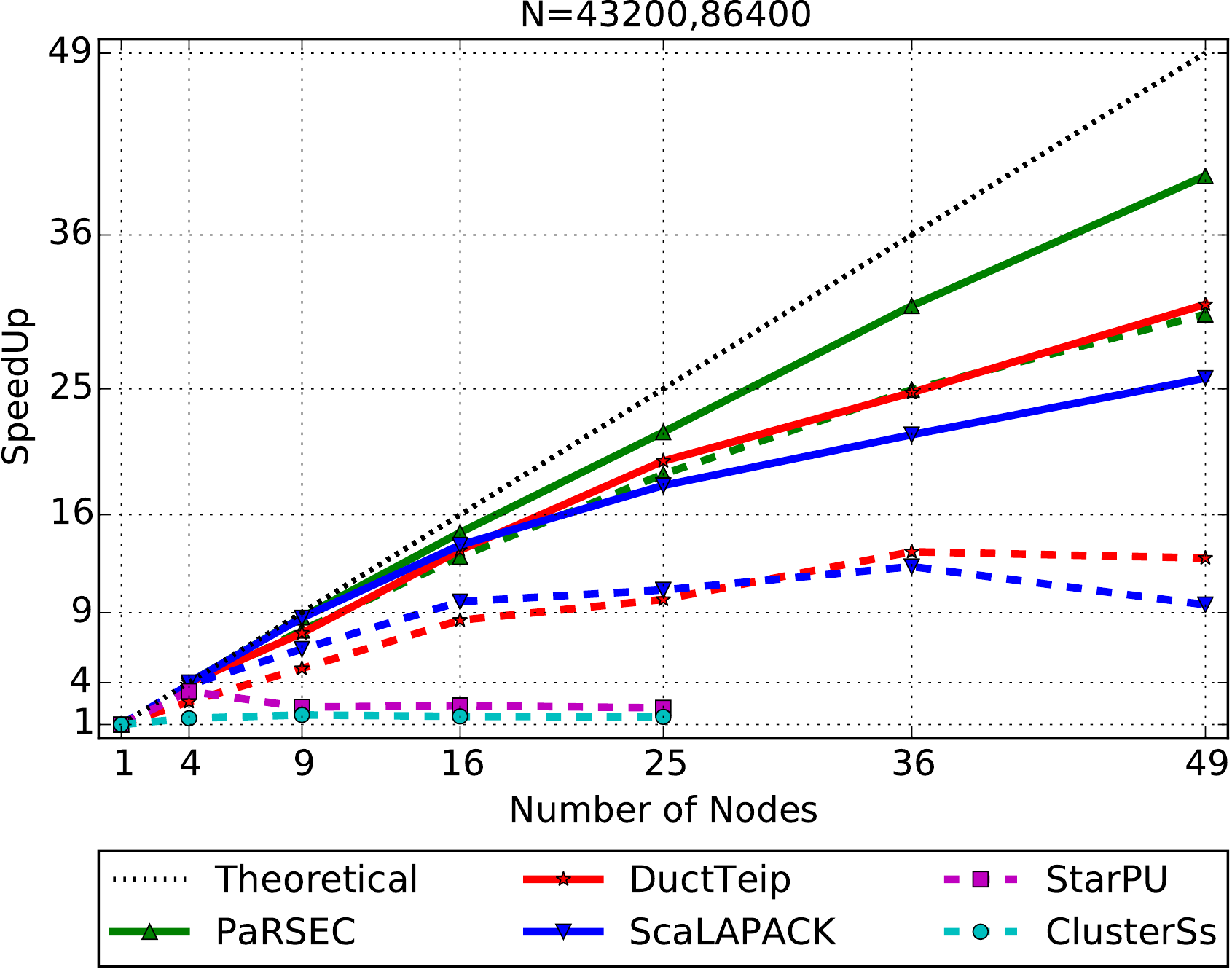}
\caption{The left subfigure shows the scaling of the different frameworks for Cholesky factorization of a matrix of size $N=43\,200$. In the right subfigure, the fixed size speedup for
problem sizes $N=43\,200$ (dashed lines) and $N=86\,400$ (solid lines) is shown.
}
\label{fig:sscale}
\end{figure}

For the strong scaling experiments, we have identified the largest problem that can be solved on a single node without performance (throughput) degradation. This corresponds to matrix size $N=43\,200$. Then, with the problem size fixed, we increase the number of nodes. Theoretically, linear scalability with the number of nodes is expected. The experimental results in terms of TFlops are shown in the left part of Figure~\ref{fig:sscale}. 

Two frameworks, PaRSEC and DuctTeip are faster than the highly optimized ScaLAPACK Cholesky factorization. Potential reasons for this could be that the ScaLAPACK implementation is a pure MPI implementation, and that the matrix storage is not tiled. Furthermore, a dynamical schedule without global barriers can be more efficient in exploiting the variabilities of modern hardware than a static schedule with global synchronization. 

PaRSEC is the fastest of all frameworks. One reason for this can be that with the embedded parametrized task graph, decisions about which task to run next can be taken quickly leading to a low overhead. DuctTeip comes out well in the comparison. A difference compared with PaRSEC is that no pre-knowledge about the data dependencies is used. We draw the conclusion that the programming model with hierarchical partitioning of data and tasks coupled with decentralized task submission and scheduling is appropriate for the architecture we are using.

Both StarPU and Cluster OmpSs perform significantly worse, especially for larger numbers of nodes. A contributing reason for StarPU can be the single level data and tasks, which increases the amount of communication, whereas in Cluster OmpSs, the centralized task submission can be an explanation for the lack of scalability. 





 
%



The right part of Figure~\ref{fig:sscale} shows the speedup for the same problem (dashed lines) for up to 49 nodes (784 cores) as well as for a problem with a larger matrix size, optimized to give maximum performance on four nodes (64 cores) (solid lines). The speedup numbers are computed as 
$S_p = T_{1}/T_p$ for the first experiment, and $S_p = 4 T_{4}/T_p$ for the second experiment. The relative performance of the frameworks is the same in both experiments, but the second experiment also shows that increasing the workload results in a reduction of the relative size of the overhead, and the speedup and scaling are thereby improved. Both StarPU and Cluster OmpSs had problems with handling large matrices and/or larger number of nodes, and therefore no results are shown for these frameworks in those cases.




\subsubsection{Weak scaling experiments}\label{sec:weakscaling}

\begin{figure}[h!]
\centering
\includegraphics[width=0.7\textwidth]{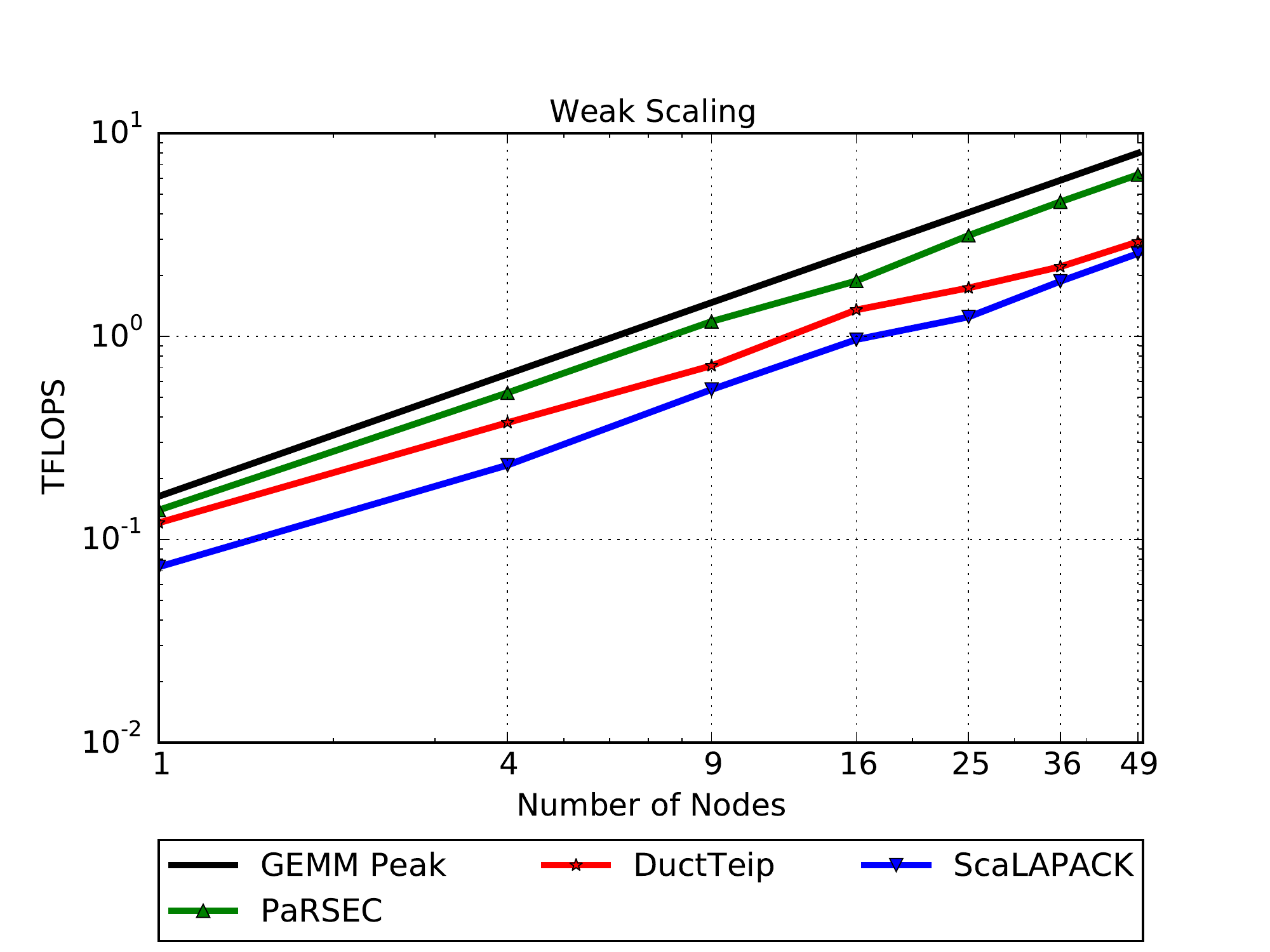}
\caption{Weak scaling scaling results for the Cholesky factorization when the problem size is scaled up from $N=10\,800$ one node such that $N^2/p$, where $p$ is the number of processes, is held constant.}
\label{fig:wscale}
\end{figure}

For the weak scaling experiments, we increase the problem size such that the communication size is constant as we also did in Section~\ref{sec:weakchol}. The problem sizes are given by $N=10\,800\sqrt{p}$ leading to to $N=75\,600$ on 49 nodes (784 cores).
Figure~\ref{fig:wscale} shows the experimental result. In absolute numbers, PaRSEC shows the best performance, followed by \ductteipw, and then ScaLAPACK. However, looking at the slope of the scaling results, PaRSEC scales almost perfectly, while \ductteip scales slightly worse than ScaLAPACK. 




\section{Conclusions}\label{sec:conc} 

We have introduced a dependency-aware hierarchical task-based programming model, where the execution order of the tasks is controlled by a versioning system. 
The data versioning approach has previously been shown to be very effective in a shared memory setting~\cite{Tillenius15}, and here we have shown that it 
is also very useful in a distributed context.
In particular, the consistency of the hierarchical task submission is automatically handled by the framework, and tasks at all levels can be executed in parallel. 
The hierarchical task decomposition helps to pace the submission of tasks such that task overflow is avoided. It also gives a lot of flexibility in choosing appropriate task sizes for the different levels. There are however a number of trade-offs that influence the performance that make it hard to select the blocking parameters. The simulation mode of DuctTeip gives some guidance regarding load balancing over nodes and communication volumes. 
%
%
%
Furthermore, we have proposed an abstract software construct, the listener, which enables the run-time system to control the communication of data. Without this abstraction level, the run-time system would not be able to detect multiple requests of the same data from a single requester or resolve the arrival of multiple versions of the same data by a single receiver. 
%

%

In the current \ductteip implementation, the programmer needs to supply the task definitions and algorithms at each level. This is an unnecessary obstacle to the ease of programming in the framework. Therefore, in an ongoing project, we are developing a unified interface, such that the algorithm and the task kernels only need to be defined once and then can be reused for all levels~\cite{Zafari17}.

A potential obstacle to performance is load imbalance between the computational nodes. The task ownership is an implicit result of the data distribution scheme, the block sizes, the process grid, and the algorithm. Clearly, these factors together may result in load imbalance. In the \ductteip framework both tasks and data are designed to be movable objects. Hence, the framework is prepared for distributed dynamical load balancing. In another ongoing project, we are investigating whether distributed load balancing can be implemented within the framework in such a way that overall performance is improved.

We have, to our knowledge, performed the first comparison involving several of the established distributed task parallel frameworks. The experiments show that \ductteip compares very well with similar frameworks and libraries regarding both absolute performance and relative speedup. Only PaRSEC, which is extremely scalable, but also encodes more information about the task graph in the compiled program, performs better.

\section*{Acknowledgments}

The computations were performed on resources provided by SNIC through Uppsala Multidisciplinary Center for Advanced Computational Science (UPPMAX) under Project p2009014.
Lennart Karlsson at UPPMAX is acknowledged for assistance concerning technical and implementational aspects in making the code run on the UPPMAX resources.

Javier Bueno Hedo, Barcelona Supercomputing Centre, Barcelona, Spain, is acknowledged for assistance with the installation of the Cluster OmpSs compilers and for providing the Cholesky factorization code for Cluster OmpSs.

Finally, we thank Igor Tominec, Uppsala University for constructive feedback on the text.


\end{document}